\documentclass[final,3p,times]{elsarticle}

\usepackage{amsgen,amsmath,amstext,amsbsy,amsopn,amssymb,subfigure}
\usepackage[linesnumbered,ruled,vlined]{algorithm2e}
\usepackage{enumitem}
\usepackage{comment}
\usepackage{booktabs}
\usepackage{xspace} 
\usepackage{graphicx}
\usepackage{subfigure}
\usepackage{epsfig}
\usepackage[labelfont=bf, labelsep=period]{caption}
\usepackage[hyphens]{url}
\usepackage{tabularx}
\usepackage[table]{xcolor}

\usepackage{setspace}

\journal{Journal of Biomedical Informatics}

\begin{document}
\onehalfspacing

\begin{frontmatter}

\title{Soft Phenotyping for Sepsis via EHR Time-aware Soft Clustering}

\author[inst1,+]{Shiyi Jiang, MS}
\author[inst2,+]{Xin Gai, BS}
\author[inst3,-]{Miriam M. Treggiari, MD, PhD}
\author[inst4]{William W. Stead, MD}
\author[inst5]{Yuankang Zhao, MS}
\author[inst5]{C. David Page, PhD}
\author[inst5,inst6,-]{Anru R. Zhang, PhD\corref{corrauthor}}

\cortext[corrauthor]{Corresponding author}
\ead{anru.zhang@duke.edu}

\affiliation[inst1]{organization={Department of Electrical \& Computer Engineering, Duke University},
                    city={Durham},
                    postcode={27708},
                    state={NC},
                    country={USA}}
\affiliation[inst2]{organization={Department of Statistical Science, Duke University},
                    city={Durham},
                    postcode={27708},
                    state={NC},
                    country={USA}}
\affiliation[inst3]{organization={Department of Anesthesiology, Duke University},
                    city={Durham},
                    postcode={27708},
                    state={NC},
                    country={USA}}
\affiliation[inst4]{organization={Department of Biomedical Informatics, Vanderbilt University},
                    city={Nashville},
                    postcode={37235},
                    state={TN},
                    country={USA}}
\affiliation[inst5]{organization={Department of Biostatistics \& Bioinformatics, Duke University},
                    city={Durham},
                    postcode={27708},
                    state={NC},
                    country={USA}}
\affiliation[inst6]{organization={Department of Computer Science, Duke University},
                    city={Durham},
                    postcode={27708},
                    state={NC},
                    country={USA}}
                    
\affiliation[+]{These authors contributed equally to this work.}
\affiliation[-]{These authors contributed equally to this work.}



\begin{abstract}
\textbf{Objective:} Sepsis is one of the most serious hospital conditions associated with high mortality. Sepsis is the result of a dysregulated immune response to infection that can lead to multiple organ dysfunction and death. Due to the wide variability in the causes of sepsis, clinical presentation, and the recovery trajectories, identifying sepsis sub-phenotypes is crucial to advance our understanding of sepsis characterization, to choose targeted treatments and optimal timing of interventions, and to improve prognostication. Prior studies have described different sub-phenotypes of sepsis using organ-specific characteristics. These studies applied clustering algorithms to electronic health records (EHRs) to identify disease sub-phenotypes. However, prior approaches did not capture temporal information and made uncertain assumptions about the relationships among the sub-phenotypes for clustering procedures.

\noindent\textbf{Methods:} We developed a time-aware soft clustering algorithm guided by clinical variables to identify sepsis sub-phenotypes using data available in the EHR.

\noindent\textbf{Results:} We identified six novel sepsis hybrid sub-phenotypes and evaluated them for medical plausibility. In addition, we built an early-warning sepsis prediction model using logistic regression.

\noindent\textbf{Conclusion:} Our results suggest that these novel sepsis hybrid sub-phenotypes are promising to provide more accurate information on sepsis-related organ dysfunction and sepsis recovery trajectories which can be important to inform management decisions and sepsis prognosis.
\end{abstract}

\begin{keyword}
Sepsis Sub-phenotyping \sep EHR \sep Soft Clustering \sep Semi-supervised Learning
\end{keyword}

\end{frontmatter}

\clearpage

\newpage
\section{\textbf{INTRODUCTION}}\label{sec:intro}
Sepsis is a life-threatening organ dysfunction syndrome secondary to a dysregulated host response to infection, and the primary cause of death from infection, especially if not recognized and treated promptly \cite{Singer2016TheThirdIC}. A hallmark of sepsis is the heterogeneity of its presentation and its prognosis, due to the variability in pathogen and immune host response interactions.

In 2016, a consensus conference provided an updated definition of sepsis, with septic shock representing a subset of sepsis in which particularly profound circulatory, cellular, and metabolic abnormalities lead to substantially increased mortality \cite{Singer2016TheThirdIC}.

The consensus definition emphasized the importance of timely recognition and prompt management of sepsis \cite{Hotchkiss2016SepsisAS}. Available therapies and management for patients with sepsis remain limited to source control, administration of antibiotics, and supportive care \cite{DeMerle2021SepsisSA}. Accumulated evidence suggests that the intrinsic heterogeneity of sepsis and variable stage at presentation posed challenges not only to clinical care but also to the conduct of clinical trials assessing interventions for sepsis. Therefore, identifying its sub-phenotypes is crucial for informing prognostic assessment and developing and evaluating effective treatment plans.

A prior study identified sepsis phenotypes at the time of patient presentation to the emergency department, using only routinely available Electronic Health Record (EHR) data in the clustering models \cite{Seymour2019DerivationVA}. The phenotypes were derived from a large observational cohort to ensure generalizability. This important study, however, did not account for the temporal registration and the rapidly evolving changes in patient physiological and laboratory values. Information acquired in the early course of sepsis can substantially enrich the clinical phenotypes, enable the identification of sub-phenotypes, and increase prognostic accuracy. Other studies have captured the dynamic nature of the clinical course in patients with sepsis using the change in the Sequential Organ Failure Assessment (SOFA) score that assesses the severity of organ dysfunction in ICU patients \cite{Schertz2023SepsisPM}. However, these scores have been used primarily as outcome measures to evaluate the overall course of organ dysfunction and to predict mortality.

To further advance the classification of sepsis, and identify potential subgroups, we incorporated medical context and temporal biomarker characteristics into the sepsis classification algorithms, early after sepsis onset.

Researchers have been studying disease phenotyping with the help of machine learning techniques and Electronic Health Records (EHRs) \cite{Afshar2019SubtypesIP, Maurits2022AFF, Zhao2019DetectingTP, Mullin2021LongitudinalKA,xu2022sepsis}, which contain large amounts of patient-level information, including demographics, vital signals, lab tests, medications, and diagnosis. However, in recent review papers, Yang et al. and He et al. \cite{Yang2022MachineLA, He2023TrendsAO} pointed out that most existing literature used purely data-driven approaches and seldom considered real-world medical use cases and corresponding medical interpretations. Limited work considers temporal information in the EHR longitudinal data. In addition, few existing studies perform non-overlapping clustering, i.e., each patient is commonly assigned to only one group (sub-phenotype).

Sepsis may initially be associated with dysfunction of one organ system and progress to involve multiple organ systems. Because of the involvement of multiple systems, a patient may exhibit more than one sub-phenotype.
We thus develop a soft clustering method that allows each patient to be assigned to more than one sub-phenotype. At the same time, we take biomarker temporal information into account and incorporate clinical information into the soft clustering algorithm. By applying transformations to the soft clustering results, we obtain six novel sepsis hybrid sub-phenotypes. We evaluate the plausibility of the results by providing a biological explanation. Additionally, built upon the soft clustering results, we train and validate a sepsis early-warning model to predict the novel sepsis hybrid sub-phenotypes. The results suggest the newly identified hybrid sub-phenotypes provide characterizations of different sepsis progressions.

\section{\textbf{BACKGROUND AND SIGNIFICANCE}}
\subsection{\textnormal{\textbf{Disease sub-phenotyping using EHR}}}
With the growing resource of EHR availability, researchers began to identify disease sub-phenotypes using EHR to better characterize the diseases and provide insights for subsequent treatment plans. Wang et al. \cite{Wang2019UnsupervisedML} proposed an algorithm that is built upon Latent Dirichlet Allocation for topic modeling to identify latent patient subgroups from three patient cohorts. Ibrahim et al. \cite{Ibrahim2019OnCS} utilized hierarchical clustering to identify sepsis sub-populations. Oh et al. \cite{Oh2022UsingSC} applied agglomerative hierarchical clustering to identify COVID-19 sub-phenotypes. Seymour et al. \cite{Seymour2019DerivationVA} discovered four novel sepsis clinical phenotypes by applying consensus K-Means clustering. However, none of the prior work utilizes the temporal information contained in the EHR. They typically used representative values at a certain time point within a defined time range, failing to capture changes in feature patterns through time.

There is a limited amount of work that considers temporal information. For instance, Xu et al. \cite{Xu2019IdentifyingSO} transformed acute kidney injury (AKI) EHR longitudinal data into vector representations using memory networks and performed K-Means clustering after applying dimensional reduction to the transformed data. They identified three novel AKI sub-phenotypes with distinct characteristics. Lasko and Mesa \cite{Lasko2019ComputationalPD} transformed longitudinal EHR data into continuous space and applied independent components analysis to identify sub-phenotypes of liver diseases. Smith et al. \cite{Smith2022OnlineCD} proposed an algorithm to detect sepsis patients using longitudinal EHR data via Jensen-Shannon Divergence. Estiri et al. \cite{Estiri2021HighthroughputPW} transformed medication and diagnosis records into vectors and performed semi-supervised learning for phenotyping. Lee and Schaar \cite{Lee2020TemporalPU} developed a dynamic clustering algorithm using deep learning for phenotyping. However, they utilized data-driven approaches and did not incorporate medical context into the designed models. We thus propose a soft clustering algorithm integrated with medical context to better characterize disease sub-phenotypes.

\subsection{\textnormal{\textbf{Soft clustering algorithms}}}
Clustering is an important group of unsupervised learning algorithms that groups data samples based on similarity with a wide range of applications, such as biomedical data analysis, anomaly detection, and building recommendation systems \cite{Xu2005SurveyOC}. Conventional clustering algorithms, such as K-Means clustering \cite{MacQueen1967SomeMF}, hierarchical clustering \cite{Johnson1967HierarchicalCS} and DBSCAN \cite{Ester1996ADA}, assign one sample to exclusively one cluster. Such algorithms are termed hard clustering.

Correspondingly, another category of algorithms that allows one sample to be assigned to multiple clusters is termed soft clustering algorithms. For instance, Fuzzy C-Means (FCM) clustering \cite{Bezdek1984FCMTF} is an algorithm that is built upon the K-Means clustering that assigns each sample with a degree of membership to each cluster. The degree of memberships to all clusters adds up to one. Cleuziou \cite{Cleuziou2008AnEV} proposed overlapping K-Means (OKM) clustering that assigns one sample to multiple clusters. Rather than expressing cluster assignment as degrees of memberships, OKM uses a set to indicate cluster assignment to each sample, where the sample is either a non-member or a member of the cluster. However, both FCM and OKM are sensitive to cluster centroid initialization and can be easily affected by outliers. Zhang et al. proposed K-Harmonic Means (KHM) clustering \cite{Zhang1999KHarmonicM} that utilizes the harmonic average to address the algorithm instability due to different cluster centroid initialization. There exist many other soft clustering methods that are introduced in the survey from Ferraro and Giordani \cite{Ferraro2020SoftC}. To the best of our knowledge, there is no work that utilizes soft clustering methods to tackle sepsis sub-phenotyping using EHR data.

\section{\textbf{MATERIALS AND METHODS}}

\begin{figure*}[!hbt]
  \centering
  \captionsetup{justification=centering}
  \includegraphics[width=.9\textwidth]{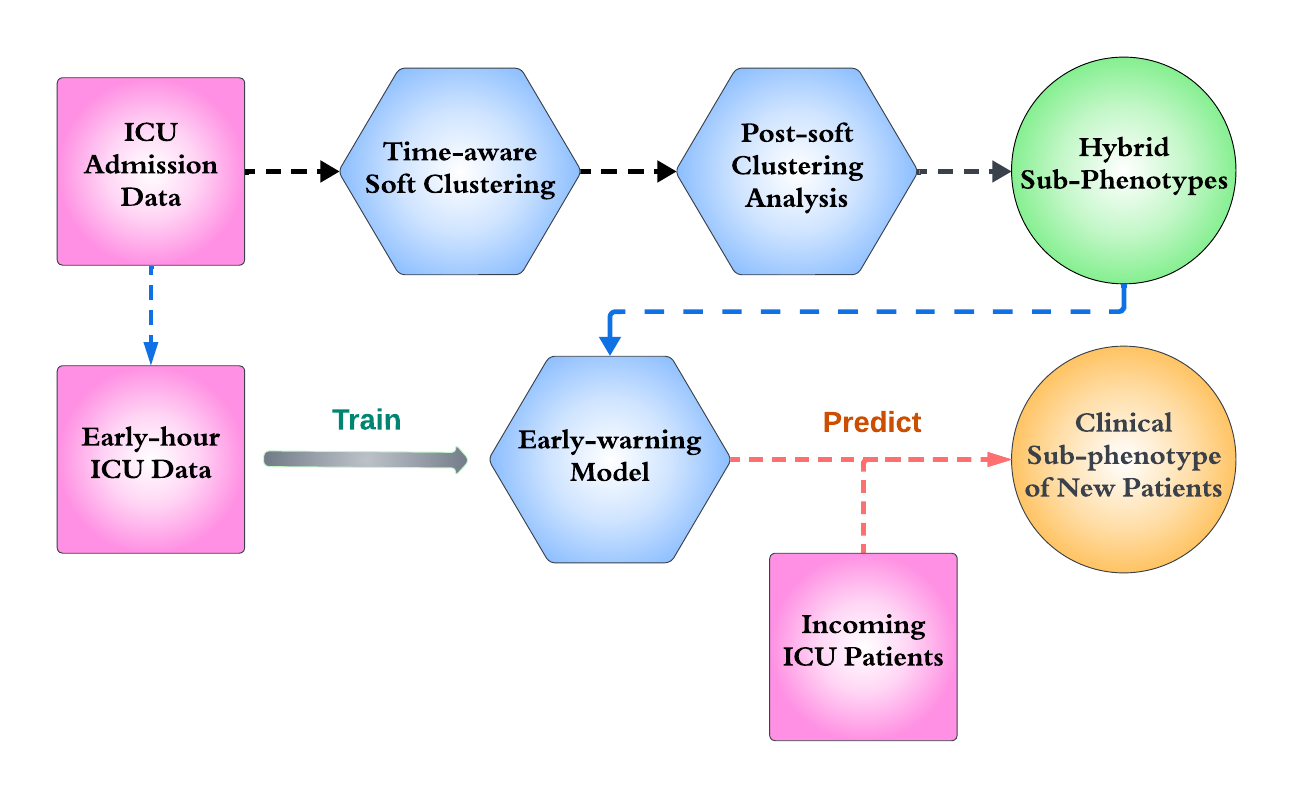}
  \caption{Overall framework of the proposed method. Pink squares indicate data, blue hexagons represent algorithms/models, and circles in green and gold describe the outcome at the training and prediction stages, respectively.}
  \label{fig:framework}
\end{figure*}
In this section, we provide a description of the datasets utilized in this study and the data selection/preprocessing steps. We next explain in detail each module of the proposed method for sepsis phenotyping illustrated in Fig. \ref{fig:framework}. We develop a time-aware soft-clustering algorithm followed by post-soft clustering analysis to identify potential novel sepsis sub-phenotypes. After careful evaluation, the resulting novel sub-phenotypes are utilized for sepsis early-warning prediction.

\subsection{\textnormal{\textbf{Data}}}
\subsubsection{\textnormal{Medical Information Mart for Intensive Care (MIMIC)-IV Database}}
The MIMIC-IV database contains de-identified medical information on over 40,000 patients admitted to the intensive care units (ICU) of the Beth Israel Deaconess Medical Center (BIDMC) from 2008 to 2019. It contains information from many aspects, such as demographics, admissions, vital signs, laboratory tests, diagnosis, and treatments. 

\subsubsection{\textnormal{eICU Collaborative Research Database}}
The eICU database contains medical records of over 200,000 patients admitted to the ICU in the continental US collected in 2014 and 2015. Similar to the MIMIC-IV database, the eICU includes information about patient demographics, admissions, diagnosis, medications, laboratory tests, etc.

\subsection{\textnormal{\textbf{Cohort selection and preprocessing}}}
We extracted patient data based on Diagnosis Related Group (DRG) Codes \cite{DRGcodes}, which are classified based on the International Classification of Diseases (ICD) diagnosis \cite{ICDcodes}, age, sex, surgical procedures, discharge status, and comorbidity. We selected patients with DRG codes 870 (septicemia or severe sepsis with mechanical ventilation $>$ 96 hours), 871 (septicemia or severe sepsis without mechanical ventilation $>$ 96 hours with major complication or comorbidity), and 872 (septicemia or severe sepsis without mechanical ventilation $>$ 96 hours without major complication or comorbidity). Since the eICU dataset does not contain DRG codes, we used the corresponding ICD codes that are mapped to the DRG codes according to \cite{DRGtoICD}. Based on the characteristics of the sepsis \cite{Otto2011TheLP}, we chose records of the first 120 hours of the last ICU stay from each patient. We chose variables included in or that contribute to the SOFA score because the third international consensus definitions for sepsis and septic shock (Sepsis-3) considers changes in SOFA as an indicator of sepsis progression \cite{Singer2016TheThirdIC}. We thus selected the following features: Arterial Blood Pressure systolic, Base Excess, Creatinine, Heart Rate, International Normalized Ratio of Prothrombin Time (PT-INR), Lactate, and Respiratory Rate.

We format the data into the form of a tensor $\mathcal{D} \in \mathbb{R}^{N \times P \times T}$, where $N$ is the number of subjects, $P$ is the number of clinical features, and $T$ is the number of time steps in hours. Since the raw data contains a large amount of missing data, we conduct missing data imputation by introducing a novel method that uses a combination of low-rank matrix completion \cite{Jain2013LowrankMC} and EHR timeline registration from our prior work \cite{Jiang23TimelineRE}. We formulate data imputation as an alternating minimization problem. The goal is to find $U \in \mathbb{R}^{N \times r}$, $V \in \mathbb{R}^{PT \times r}$, and $\tau$, such that the following objective function in Equation \eqref{impute} is minimized: 
\begin{equation}\label{impute}
    \min_{U, V, \tau} \sum_{i=1}^{N} \sum_{j=1}^{PT} A_{ij} \cdot \left(D_\tau[i,j] - U_{[i,:]} V_{[j,:]}^\top\right)^2.
\end{equation}
Here, $r$ is the matrix rank, $\tau$ is the discrete amount of time shift and $A \in \mathbb{R}^{N \times PT}$ is the indicator for the missing data. The alternating minimization contains two steps: (1) Obtain $U$ and $V$ from low-rank matrix completion on $D_{\tau} \in \mathbb{R}^{N \times PT}$ while $\tau$ is fixed. Note that $D_{\tau}$ is a matrix with $\mathcal{D}$ shifted by $\tau$ and reshaped into the dimension of $\mathbb{R}^{N \times PT}$. (2) Find optimal shift $\tau$ while $U$ and $V$ are fixed. We repeat the above steps until convergence. The final imputed data tensor $\mathcal{D}$ is approximated by $U \cdot V^T$, reshaped to a tensor of dimension $\mathbb{R}^{N \times P \times T}$.

\subsection{\textnormal{\textbf{Time-aware soft clustering}}}
We developed a time-aware soft clustering algorithm for EHR data inspired by the work of Khanmohammadi et al. \cite{Khanmohammadi2017AnIO}. It is based on a hybrid of the harmonic K-Means clustering algorithm and the overlapping K-Means clustering algorithm. This hybrid algorithm is less sensitive to the initial cluster centroids selection and thus has improved algorithm stability.

We present the proposed soft clustering algorithm in Algorithm \ref{algo:1} (see Appendix). We denote all data records from the selected cohort as $D = \{x_i, i = 1, ..., N\}$, where $x_i \in \mathbb{R}^{P \times T}$ represents data of each subject $i$ of the $N$ total subjects (equivalent to $\mathcal{D}[i,:,:]$). To represent clinical context, we denote $L$ as a collection of binary vectors $l_i \in \mathbb{R}^3$, where each element in $l_i$ indicates the existence of an organ dysfunction type on subject $i$. For sub-phenotyping, we selected organ dysfunctions representing the lung, liver, and kidney based on their contribution to the SOFA score computation. We create $L$ based on groups of ICD-9 codes for the three types of organ dysfunction: liver (570$.^{*}$-573$.^{*}$), kidney (580$.^{*}$-589$.^{*}$), and lung-related (510$.^{*}$-519$.^{*}$) diseases \cite{ICDcodes}. For instance, if subject $i$ is diagnosed with ICD-9 codes 573.9 and 584.9 but none from 510$.^{*}$ to 519$.^{*}$, the corresponding $l_i$ is $(1, 1, 0)^T$.

According to Basu et al. \cite{Basu2002SemisupervisedCB}, labeled data used to develop initial cluster centroids and cluster constraints effectively enhances the quality and stability of the clustering result. We thus initialized $K$ cluster centroids $\{temp_{k}, k=1, ..., K\}$ using the average of subjects with each of the single types of organ dysfunction (i.e., one cluster centroid for each of the liver, kidney, and lung-only dysfunction subject groups). We then perform cluster assignments to each subject according to Algorithm \ref{algo:2} (see Appendix) based on the overlapping K-Means clustering. We compute the distance between the subject $i$ and each cluster centroid and assign the subject to its nearest cluster centroid. Subsequent cluster assignments depend on $\Phi(x_i)$, the average of the assigned cluster centroids to subject $i$, and $\Phi(x_i)'$, the average of the assigned centroids and the nearest candidate centroid to subject $i$. If the subject is closer to $\Phi(x_i)'$ than to $\Phi(x_i)$, the individual is then assigned to the nearest candidate cluster centroid. After obtaining cluster assignments $\{m_i^{(0)}, i=1, ..., N\}$, where each $m_i^{(0)}$ is a set of cluster membership indicators, we update the cluster centroids by applying transformations shown in Step 2 of the Algorithm \ref{algo:1}. We iteratively update the cluster centroids and cluster assignments until convergence.

To incorporate clinical context into the algorithm, we employ semi-supervised learning that calibrates cluster centroids after updating cluster assignments at each iteration based on the ICD group information $L=\{l_i, i=1, ..., N\}$ of each subject shown in Algorithm \ref{algo:3} (see Appendix). We compute a weighted sum of distances $unsupLoss$ for all subjects which is built upon the objective function of the fuzzy C-Means algorithm \cite{Bezdek1984FCMTF}. Additionally, we compute $supLoss$ to enforce each subject with a single ICD group label to be closer to the targeted cluster centroid and further away from the non-targeted cluster centroids. Note that we assume each cluster centroid represents a designated organ dysfunction type. We use scalar hyperparameters $\beta_1$ and $\beta_2$ to adjust the strength of the constraint within $supLoss$. $\beta_1$ controls $tLoss$, the degree to which each subject’s (with a single ICD group labeled) distance to the targeted cluster centroid $temp_{l \in l_i}$. $\beta_2$ adjusts $ntLoss$, the degree to which each subject’s distance to the non-targeted cluster centroids $\{temp_{l \notin l_i}\}$. Finally, we calibrate cluster centroids by applying the stochastic gradient descent (SGD) to the $totLoss$.

After the iterative updates and calibrations of the cluster centroids reach convergence, we output a distance matrix $\{d_{ik}\} \in \mathbb{R}^{N \times K}$, where each element $d_{ik}$ indicates the distance between the subject $i$ to the cluster centroid $k$. Note that when computing the distance of each subject to a cluster centroid, we only consider the first 24 hours of data in the ICU for features including systolic blood pressure, base excess, and respiratory rate since the effects of treatments may affect subsequent data patterns for different phenotypes. The resulting distance for each above-mentioned feature was multiplied by five to ensure the computed distance for each feature having the same scale. We used entire 120-hour data to compute distance for the rest of the features. We then compute cluster membership matrix $\{\mu_{ik}\}$, where each element denotes the degree of membership of subject $i$ in relation to the cluster centroid $k$ shown in Step 5 of the Algorithm \ref{algo:1}.

\subsection{\textnormal{\textbf{Post-soft clustering analysis}}}
As described in the previous section, we now obtain a cluster membership vector $U_i = (\mu_{i1}, \mu_{i2}, ..., \mu_{iK})^T$ for each subject $i$. Considering the temporal statistical heterogeneity of the time series clinical data, i.e., two subjects with different sepsis sub-phenotypes may have opposite trends in a given time range but can still be grouped into the same cluster, we introduce an additional indicator to quantify this temporal data heterogeneity, which, in this context, pertains to similarity with the cluster centroids (sub-phenotypes). We term this indicator ABM and define it in Equation \eqref{eq:1} as follows:
\begin{equation}
\label{eq:1}
    ABM_i = 1 - \frac{min(d_{i1}, d_{i2}, ..., d_{iK})}{dist}^{\frac{1}{3}}
\end{equation}
where $dist = max(\{d_{ik}\})$, which is computed across all subjects. The value of ABM ranges from 0 to 1, and the smaller the value, the further away it is from the cluster centroids (sub-phenotypes). We hypothesize that in datasets utilized in this study, a lower ABM value indicates increased severity of the health condition, and we further explain the hypothesis in Section 5. By combining the cluster membership vector $U_i$ and the indicator ABM, we obtain a final representation of the soft clustering result $R_i$ for each subject $i$, where we denote as $R_i = (\mu_{i1}, \mu_{i2}, ..., \mu_{iK}, ABM_i)^T$. Intuitively, this representation captures the composition of clinical sub-phenotypes of each patient as a mixture of the three primary organ dysfunction phenotypes and the severity of the patient’s health condition due to the disease.

We next use the K-Medoids clustering \cite{Park2009ASA} to group all $R_i$ to identify potential sepsis hybrid sub-phenotypes for a better classification of the soft clustering result. Note that these hybrid sub-phenotypes are combinations of the cluster centroids (sub-phenotypes) from the soft clustering results. We evaluate the quality of the clustering results by computing the mean Silhouette score \cite{Rousseeuw1987SilhouettesAG} across all data samples and via clinical interpretation. The Silhouette score $s_i$ for a single data sample $i$ is computed according to Equation \eqref{eq:Silhouette}, where $a_i$ is the average distance of data sample $i$ to every other samples within the assigned cluster, and $b_i$ is the average distance of data sample $i$ to all samples in cluster that is the closest to the assigned cluster.
\begin{equation}
\label{eq:Silhouette}
    s_i = \frac{b_i - a_i}{max(a_i, b_i)}
\end{equation}

To optimize the post-soft clustering analysis using K-Medoids clustering, we computed the Silhouette scores for clustering results using different cluster numbers ranging from 2 to 20. Besides choosing the cluster number from the purely data-driven perspective, we also considered the medical interpretability of the data, i.e., the cluster number should be greater than 3, considering that the resulting clusters should be combinations of the clusters from the soft clustering.

\subsection{\textnormal{\textbf{Early-warning prediction}}}
After obtaining the results of the K-Medoids clustering, we treat the resulting cluster assignment of each subject as a ground truth label of the sepsis hybrid sub-phenotype. We utilize the first 12, 24, and 48 hours of the ICU data to predict the sepsis hybrid sub-phenotype of the subjects as an early-warning model. We follow the work of Lipton et al. \cite{Lipton2015LearningTD} to derive statistical features from the time-series data and to compute the validation metrics. Specifically, we select seven different time windows (explained in Section 4.1) for each feature of the patient and then compute the mean, standard deviation, maximum, minimum, and skewness. Logistic regression (LR) is used for sepsis early-warning prediction due to its robustness on EHR classification tasks based on our prior work \cite{Jiang23TimelineRE}.

\subsection{\textnormal{\textbf{Experimental setting}}}
We formatted all subject's data as $D \in \mathbb{R}^{N \times P \times T}$, where $N$ is the number of subjects, $P$ is the number of features, and $T$ is the number of hours recorded since the ICU admission. Normalization was applied to each of the features. We set the number of clusters $K=3$ for soft clustering, where the cluster centroids correspond to the liver, kidney, and lung dysfunction type. We selected $\eta=2,\; \beta_1=10,\; \beta_2=0.01$ and conducted the iterative clustering for 200 epochs. 
We selected $\eta=2$ in alignment with the convention in the literature \cite{Ferraro2020SoftC}, and we tested $\beta_1$ in the range of $[1 \times 10^{-2}, 10]$ and $\beta_2$ in $[1 \times 10^{-3}, 10]$. We chose the combination that yields the best Silhouette score in the post-soft clustering analysis. The SGD utilized for cluster centroids calibration was performed at a learning rate of $1 \times 10^{-5}$. The K-Medoids clustering was performed with the number of clusters equal to six. All experiments were conducted using the PyTorch library \cite{Paszke2019PyTorchAI}.

We computed features for sepsis early-warning prediction based on the work of Lipton et al. \cite{Lipton2015LearningTD}, where the maximum, minimum, mean, standard deviation, and skewness were computed for seven different time windows from each of the features: the entire feature sequence, the first 10\%/25\%/50\%, and the last 10\%/25\%/50\% of the feature sequence. Note that all features were computed from the imputed data. We applied logistic regression (LR) using the default settings for the prediction using the Scikit-learn library \cite{Pedregosa2011ScikitlearnML}.

\begin{figure*}[!hbt]
  \centering
  \captionsetup{justification=centering}
  \includegraphics[width=.95\textwidth]{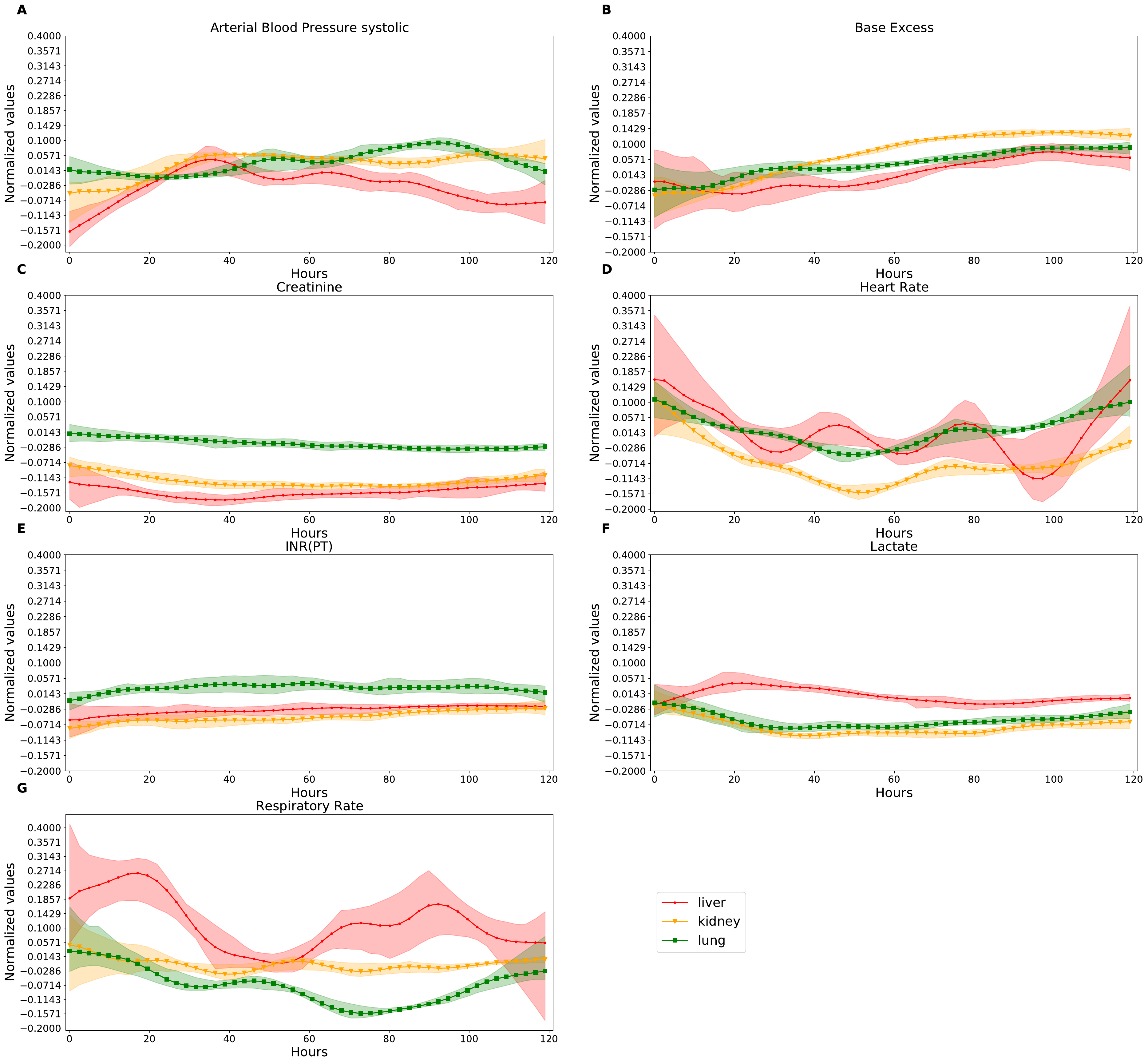}
  \caption{Soft clustering centroids per feature after smoothing obtained from the MIMIC-IV dataset. The red centroid is initialized with the liver dysfunction type; the yellow centroid with the kidney dysfunction type; and the green centroid with the lung dysfunction type.}
  \label{fig:1}
\end{figure*}

\subsection{\textnormal{\textbf{Evaluation metrics}}}
We assess the clustering results by computing the average Silhouette score ranging from -1 to 1, shown in Equation \eqref{eq:2}. The higher the score, the better the cluster quality. $a_i$ represents the mean distance between subject $i$ and other subjects within the cluster (intra-cluster distances), and $b_i$ represents the mean distance between subject $i$ and subjects from other clusters (inter-cluster distances). We compute individual Silhouette scores $s_i$ and obtain the average value.
\begin{equation}
\label{eq:2}
    s_i = \frac{b_i - a_i}{max(a_i, b_i)}
\end{equation}

In addition, we evaluate the sepsis hybrid sub-phenotype early-warning prediction using accuracy, precision, recall, and Area Under Precision-Recall Curve (AUPRC) following the work of Gao et al. \cite{Gao2020DrAC}.

\section{\textbf{RESULTS}}
\subsection{\textnormal{\textbf{Evaluation on soft clustering centroids}}}
We first present and evaluate the centroids of the three clusters obtained from the MIMIC-IV dataset shown in Fig. \ref{fig:1}, where each of them was initialized as an organ dysfunction type and was then iteratively updated to a novel sepsis sub-phenotype after reaching convergence. In a certain sense, all results of the soft clustering algorithm can be regarded as a mixture of these three sub-phenotypes. Fig. \ref{fig:1} visualizes patterns of the three cluster centroids per feature. Note that cluster centroids visualization using the eICU dataset is provided in Supplement Fig. \ref{fig:s1}.

\subsection{\textnormal{\textbf{Post-soft clustering analysis}}}
\begin{figure}[!hbt]
  \centering
  \captionsetup{justification=centering}
  \includegraphics[width=.5\linewidth]{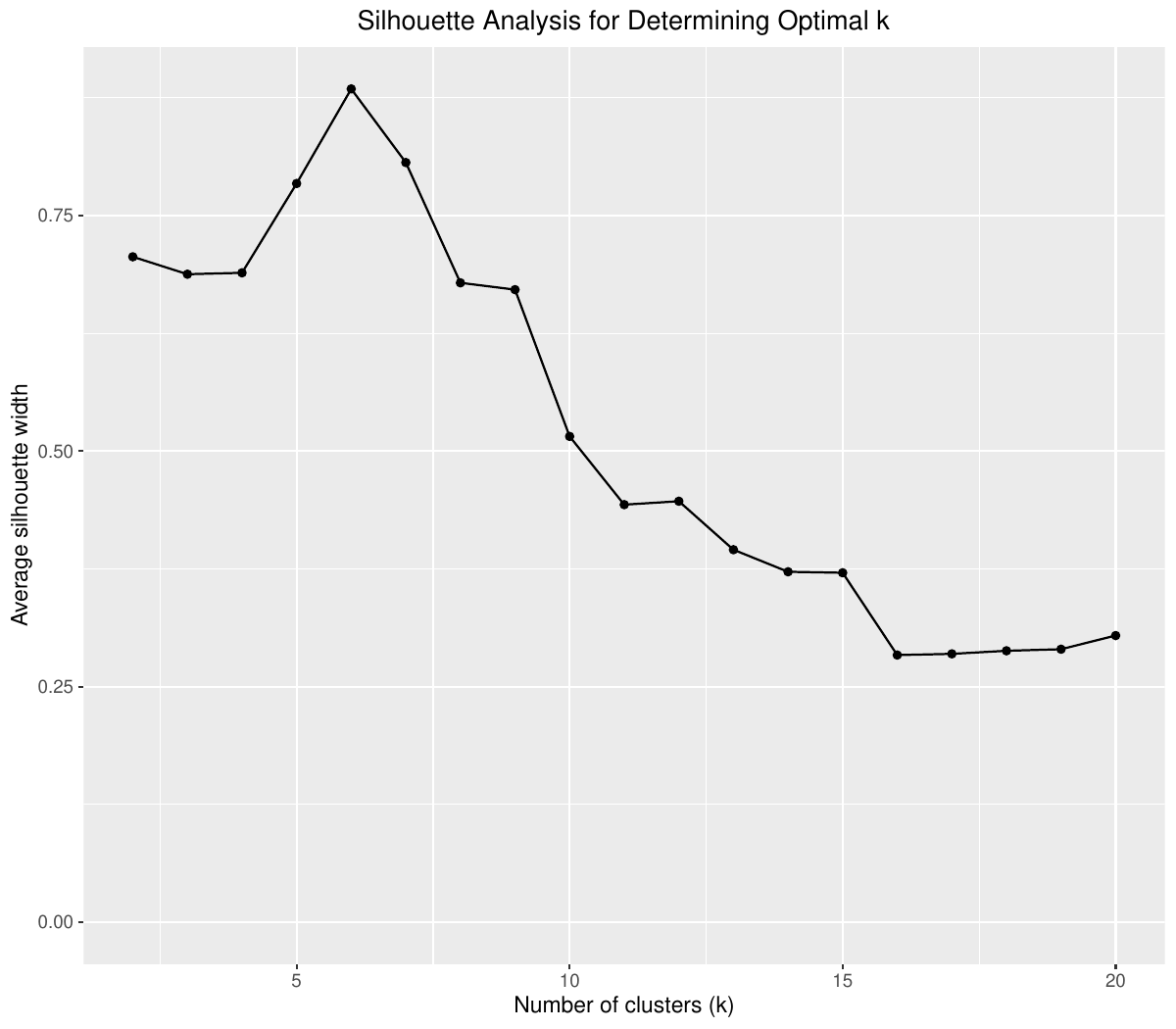}
  \caption{Cluster number selection for K-Medoids clustering using the MIMIC-IV dataset.}
  \label{fig:2}
\end{figure}

Fig. \ref{fig:2} shows the average Silhouette score computed using each cluster number. We observe that at cluster number equals 6, the cluster quality is optimal, yielding the highest Silhouette score. We thus chose 6 to be the cluster number for post-soft clustering analysis. We made the same observation using the eICU dataset from Supplement Fig. \ref{fig:s2}.

\begin{figure*}[!hbt]
  \centering
  \captionsetup{justification=centering}
  \includegraphics[width=\textwidth]{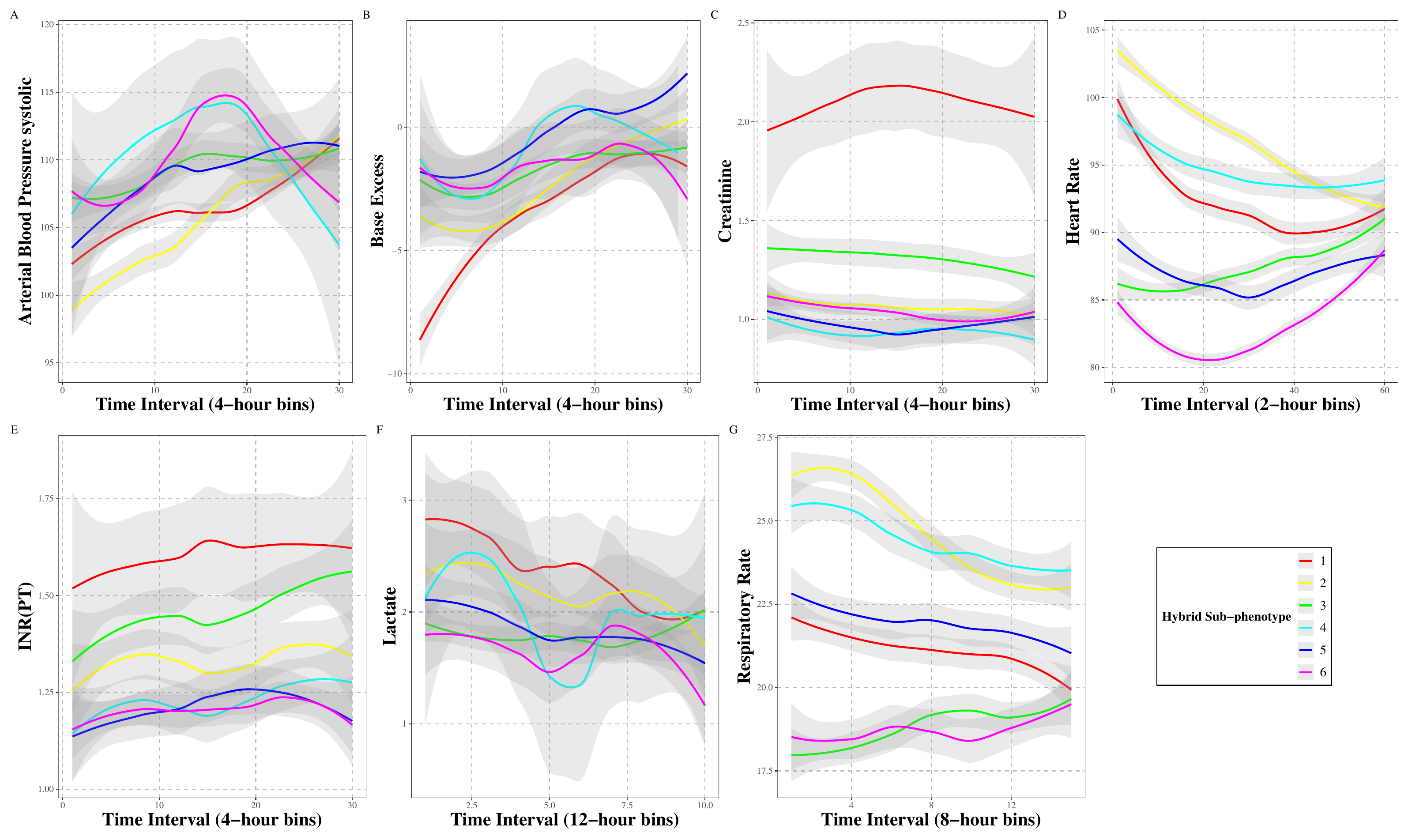}
  \caption{Comparisons of feature values between sepsis hybrid sub-phenotypes using the MIMIC-IV dataset.}
  \label{fig:3}
\end{figure*}

\begin{figure*}[!hbt]
  \centering
  \captionsetup{justification=centering}
  \includegraphics[width=\textwidth]{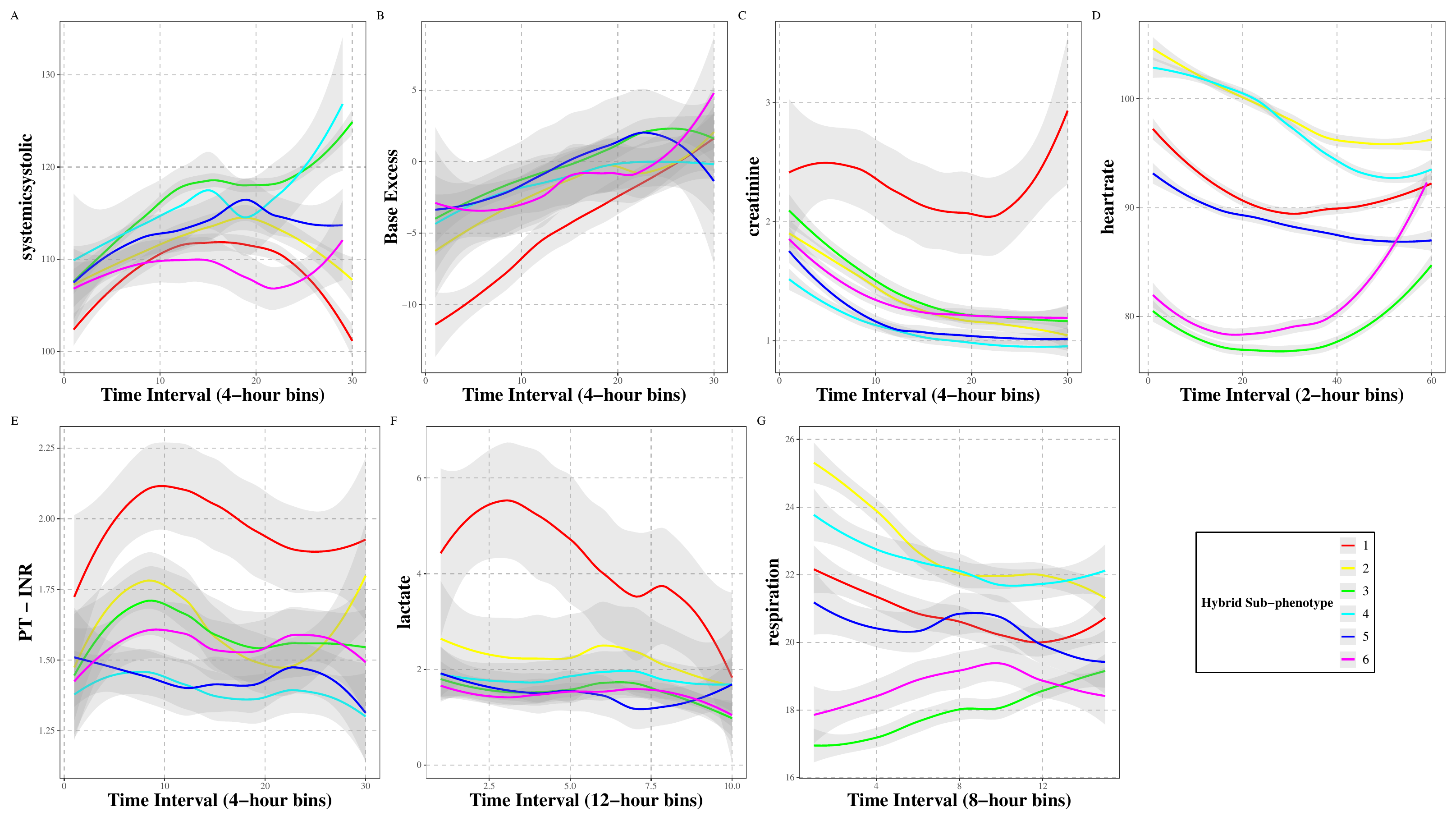}
  \caption{Comparisons of feature values between sepsis hybrid sub-phenotypes using the eICU dataset.}
  \label{fig:4}
\end{figure*}

We present in Table \ref{tab:1} the resulting 6 cluster centroids obtained from the MIMIC-IV and the eICU datasets, respectively, which indicate 6 potential sepsis hybrid sub-phenotypes. We rank sepsis hybrid sub-phenotypes based on the severity of patient health conditions from the most to the least severe as indicated by the ABM value. We also show the median and the interquartile range (IQR) of the clusters per feature in Fig. \ref{fig:3} and Fig. \ref{fig:4}. Separate figures of feature values per hybrid sub-phenotype using the MIMIC-IV and the eICU datasets are provided in the Supplement.

\begin{table}[!hbt]
\centering
\caption{K-Medoids cluster (hybrid sub-phenotype) centroids obtained from the MIMIC-IV dataset (left) and eICU dataset (right).}
\label{tab:1}
\resizebox{.44\textwidth}{!}{%
\begin{tabular}{ccccc}
\toprule
MIMIC-IV & $\mu_1$ (liver) & $\mu_2$ (kidney) & $\mu_3$ (lung) & ABM \\ \midrule
Hybrid Sub-phenotype 1 & 0.33 & 0.32 & 0.35 & 0.47 \\
Hybrid Sub-phenotype 2 & 0.44 & 0.28 & 0.28 & 0.58 \\
Hybrid Sub-phenotype 3 & 0.24 & 0.37 & 0.39 & 0.61 \\
Hybrid Sub-phenotype 4 & 0.55 & 0.24 & 0.21 & 0.70 \\
Hybrid Sub-phenotype 5 & 0.36 & 0.35 & 0.29 & 0.71 \\
Hybrid Sub-phenotype 6 & 0.18 & 0.46 & 0.36 & 0.71 \\ \bottomrule
\end{tabular}%
}
\resizebox{.44\textwidth}{!}{%
\begin{tabular}{ccccc}
\toprule
eICU & $\mu_1$ (liver) & $\mu_2$ (kidney) & $\mu_3$ (lung) & ABM \\ \midrule
Hybrid Sub-phenotype 1 & 0.33 & 0.34 & 0.33 & 0.36 \\
Hybrid Sub-phenotype 2 & 0.4 & 0.32 & 0.28 & 0.49 \\
Hybrid Sub-phenotype 3 & 0.24 & 0.35 & 0.41 & 0.51 \\
Hybrid Sub-phenotype 4 & 0.48 & 0.3 & 0.22 & 0.63 \\
Hybrid Sub-phenotype 5 & 0.31 & 0.35 & 0.34 & 0.63 \\
Hybrid Sub-phenotype 6 & 0.17 & 0.37 & 0.46 & 0.66 \\ \bottomrule
\end{tabular}%
}
\end{table}

\begin{table}[hbt!]
\centering
\caption{Mortality rate of each sepsis hybrid sub-phenotype group from the MIMIC-IV and the eICU datasets.}
\label{tab:5}
\resizebox{.44\textwidth}{!}{%
\begin{tabular}{@{}ccc@{}}
\toprule
 & MIMIC-IV Mortality (\%) & eICU Mortality (\%) \\ \midrule
Hybrid Sub-phenotype 1 & 78.71 & 42.73 \\
Hybrid Sub-phenotype 2 & 75.77 & 26.51 \\
Hybrid Sub-phenotype 3 & 71.78 & 12.45 \\
Hybrid Sub-phenotype 4 & 53.1 & 7.39 \\
Hybrid Sub-phenotype 5 & 56.15 & 13.74 \\
Hybrid Sub-phenotype 6 & 55.02 & 7.32 \\ \bottomrule
\end{tabular}%
}
\end{table}

In addition, we further summarize the patient outcome of each hybrid sub-phenotype group in terms of mortality rate in Table \ref{tab:5} to evaluate the discovered hybrid sub-phenotypes.

\subsection{\textnormal{\textbf{Early-warning prediction}}}
As mentioned in Section 3.5, we developed a sepsis hybrid sub-phenotype early-warning prediction model using the post-soft clustering results as ground truth labels. We present the results in terms of accuracy, precision, recall, and AUPRC in Table \ref{tab:2} and Table \ref{tab:4} obtained from the MIMIC-IV and the eICU datasets, respectively. We compare the results with the result of using the whole 120-hour ICU data.

\begin{table}[!hbt]
\centering
\caption{Early-warning prediction results using the MIMIC-IV dataset. Precision, recall, and AUPRC are computed by the average of the "one-vs-rest" setting.}
\label{tab:2}
\resizebox{.44\textwidth}{!}{%
\begin{tabular}{ccccc}
\toprule
Hours & Precision & Recall & Accuracy & AUPRC \\ \midrule
12 & 0.609 & 0.601 & 0.621 & 0.618 \\
24 & 0.661 & 0.65 & 0.668 & 0.671 \\
48 & 0.598 & 0.587 & 0.61 & 0.615 \\
120 & 0.546 & 0.519 & 0.552 & 0.54 \\ \bottomrule
\end{tabular}%
}
\end{table}

\begin{table}[!hbt]
\centering
\caption{Early-warning prediction results using the eICU dataset. Precision, recall, and AUPRC are computed by the average of the "one-vs-rest" setting.}
\label{tab:4}
\resizebox{.44\textwidth}{!}{%
\begin{tabular}{ccccc}
\toprule
Hours & Precision & Recall & Accuracy & AUPRC \\ \midrule
12 & 0.505 & 0.509 & 0.524 & 0.508 \\
24 & 0.569 & 0.565 & 0.576 & 0.565 \\
48 & 0.557 & 0.558 & 0.567 & 0.565 \\
120 & 0.531 & 0.531 & 0.54 & 0.547 \\ \bottomrule
\end{tabular}%
}
\end{table}

\section{\textbf{DISCUSSION}}
As can be seen from Fig. \ref{fig:1}, the three sub-phenotypes are different from each other, suggesting that the proposed semi-supervised soft clustering algorithm can generate a clear separation between the clusters. The cluster centroid in red features an elevated lactate level and a low base excess level. The cluster centroid in green exhibits elevated creatinine and INR(PT) levels. We observe that the characteristics of each of the cluster centroids do not necessarily match their patterns from the original initialization.

We observe that patients in hybrid sub-phenotype 1 obtained from both datasets exhibit the most severe health condition indicated by the lowest ABM value. The corresponding even degrees of membership to primary sub-phenotypes 1-3 ($\mu_1$-$\mu_3$ in Table \ref{tab:1}) suggest that subjects in the group may experience multiple organ failures, reflected by the lowest base excess level, the highest creatinine, INR(PT), and lactate levels compared to patients in other hybrid sub-phenotype groups.

Patients in hybrid sub-phenotypes 2 and 3 obtained from both datasets have moderate-severity health conditions. Hybrid sub-phenotype 2 subjects from both datasets have a higher degree of membership to sub-phenotype 1 ($\mu_1$), suggesting that the subjects align more closely with liver-related characteristics, reflected by moderately high levels of lactate and INR (PT). Hybrid sub-phenotype 3 subjects from the MIMIC-IV dataset have high degrees of membership to sub-phenotypes 2 and 3 ($\mu_2$ and $\mu_3$), with characteristics aligned more with kidney and lung-related diseases, reflected by the moderately high creatinine level. However, we do not observe an abnormal respiratory rate. Hybrid sub-phenotype 3 subjects in the eICU dataset feature lung-related dysfunction indicated by the high degree of membership to sub-phenotype 3 ($\mu_3$). Similarly, we do not observe an abnormal respiratory rate in the group.

Patients in hybrid sub-phenotypes 4, 5, and 6 obtained from both datasets experience a health condition of less severity. Hybrid sub-phenotype 4 subjects yield a high degree of membership to sub-phenotype 1 ($\mu_1$), reflected by an elevated lactate level, which is consistent across both datasets.

Hybrid sub-phenotype 5 subjects from the MIMIC-IV dataset align more closely with sub-phenotypes 1 and 2 ($\mu_1$ and $\mu_2$), reflected by slightly elevated lactate and creatinine levels. Hybrid sub-phenotype 5 subjects from the eICU dataset do not show significant organ dysfunction, implied by the combination of even degrees of membership to all three types of primary sub-phenotypes and a higher ABM value. Hybrid sub-phenotype 6 subjects from the MIMIC-IV dataset obtain a high degree of membership to sub-phenotype 2 ($\mu_2$), with a slightly elevated creatinine level. Hybrid sub-phenotype 6 subjects from the eICU dataset may exhibit lung-related dysfunction implied by the high degree of membership to sub-phenotype 3 ($\mu_3$). However, we do not observe an abnormal respiratory rate.

We find consistent patient characteristics of the hybrid sub-phenotypes 1, 2, and 4 across the MIMIC-IV and the eICU datasets. Discrepancies in hybrid sub-phenotypes 3, 5, and 6 can be attributed to the heterogeneity of patient cohorts between the MIMIC-IV and the eICU datasets. The similar findings across two distinct patient cohorts suggest that our proposed method could provide medically meaningful sepsis sub-phenotypes if given consistent datasets, which will be assessed in our future work. 

We observe that hybrid sub-phenotype 1 with subjects highly associated with all three organ dysfunctions has the lowest ABM, and we notice patients in a cluster with a lower ABM, such as hybrid sub-phenotype 2, yield more abnormal feature values (e.g., creatinine and lactate levels) compared to patients in a cluster with a higher ABM, such as hybrid sub-phenotype 4. We thus hypothesize a potential association between the ABM value that is derived from a data-driven perspective with the severity of the patient’s health condition based on feature values. Further exploration with other hospital data is necessary to test this hypothesis.

We observe that in the MIMIC-IV dataset, hybrid sub-phenotype 1 group exhibits the highest mortality; hybrid sub-phenotypes 2 and 3 have relatively lower mortality; hybrid sub-phenotypes 4-6 yield the lowest mortality among all the groups. This observation in mortality aligns with the ABM indicator shown in Table \ref{tab:1} that a lower ABM value corresponds to a higher mortality of the group. In the eICU dataset, the hybrid sub-phenotype 1 group shows the highest mortality; patients in the hybrid sub-phenotype 2 have moderate mortality; hybrid sub-phenotypes 3 and 5 exhibit low mortality; hybrid sub-phenotypes 4 and 6 yield the lowest mortality. We notice that the ABM values of hybrid sub-phenotypes 3 and 5 in the eICU dataset do not align with mortality well.

We obtain mixed results using early-hour ICU data for sepsis hybrid sub-phenotype prediction. The best prediction performance occurs when using the first 24-hour ICU data with an accuracy of 0.668 and an AUPRC of 0.671 using the MIMIC-IV dataset. Similarly, we obtain the best performance using the first 24-hour ICU data with an accuracy of 0.576 and an AUPRC of 0.565 using the eICU dataset. 

The current study has some limitations. We did not include cardiovascular dysfunction as part of the clinical context of the proposed algorithm because this would also involve incorporating treatment information on vasopressor use. We plan to further evaluate the cardiovascular component as we further extend and integrate treatment information into our model. Additionally, to prove the generalizability of the proposed method, further evaluations need to be done on databases utilizing coding systems other than the ICD codes. Other improvements to consider include validation of the proposed method using private hospital data and consideration of the changes in the assigned sub-phenotypes over time. Note that phenotyping on non-ICU patients is out of the scope of this study given the wider time range and higher sparsity of the records compared to ICU data. Different approaches targeting non-ICU data will be investigated in our future work.

\section{\textbf{CONCLUSION}}
Sepsis sub-phenotyping is a crucial but complex area of research. To advance the classification of sepsis sub-phenotypes and incorporate temporal changes over time, we proposed a novel soft clustering algorithm that incorporates temporal and medical context using EHR data. Our results suggest the newly discovered six hybrid sub-phenotypes are medically plausible. The sepsis early-warning prediction model we created that builds upon our sub-phenotyping findings yields promising results.

\section*{\textbf{AUTHOR CONTRIBUTIONS}}
Methodology, S.J., X.G., A.Z.; software, S.J., X.G., Y.Z.; writing--original draft preparation, S.J.; writing--review and editing, S.J., X.G., M.T., W.S., D.P., A.Z.; All authors have read and agreed to the published version of the manuscript.

\section*{\textbf{COMPETING INTERESTS}}
All authors declare no financial or non-financial competing interests.

\section*{\textbf{DATA AVAILABILITY}}
The datasets generated and/or analyzed during the current study are available in the MIMIC repository, \url{https://mimic.mit.edu}, and the eICU Collaborative Research Database, \url{https://eicu-crd.mit.edu}.

\section*{\textbf{CODE AVAILABILITY}}
The underlying code for this study is publicly available at: \url{https://github.com/Shiyi-J/EHR_Sepsis_Soft_Phenotyping}.

\bibliographystyle{vancouver}
\bibliography{references}

\begin{thebibliography}{10}

\bibitem{Singer2016TheThirdIC}
Singer M, et~al.
\newblock {The Third International Consensus Definitions for Sepsis and Septic
  Shock (Sepsis-3)}.
\newblock JAMA. 2016;315(8):801--810.

\bibitem{Hotchkiss2016SepsisAS}
Hotchkiss RS, Moldawer LL, Opal SM, Reinhart K, Turnbull IR, Vincent JL.
\newblock {Sepsis and septic shock}.
\newblock Nature Reviews Disease Primers. 2016;2:16045.

\bibitem{DeMerle2021SepsisSA}
DeMerle KM, et~al.
\newblock {Sepsis Subclasses: A Framework for Development and Interpretation}.
\newblock Critical care medicine. 2021;49(5):748--759.

\bibitem{Seymour2019DerivationVA}
Seymour CW, et~al.
\newblock {Derivation, Validation, and Potential Treatment Implications of
  Novel Clinical Phenotypes for Sepsis.}
\newblock JAMA. 2019;321(20):2003--2017.

\bibitem{Schertz2023SepsisPM}
Schertz AR, Lenoir KM, Bertoni AG, Levine BJ, Mongraw-Chaffin M, Thomas KW.
\newblock {Sepsis Prediction Model for Determining Sepsis vs SIRS, qSOFA, and
  SOFA}.
\newblock JAMA Network Open. 2023;6(8):e2329729.

\bibitem{Afshar2019SubtypesIP}
Afshar M, et~al.
\newblock {Subtypes in patients with opioid misuse: A prognostic enrichment
  strategy using electronic health record data in hospitalized patients}.
\newblock PLoS ONE. 2019;14(7):e0219717.

\bibitem{Maurits2022AFF}
Maurits MP, et~al.
\newblock {A framework for employing longitudinally collected multicenter
  electronic health records to stratify heterogeneous patient populations on
  disease history}.
\newblock Journal of the American Medical Informatics Association: JAMIA.
  2022;29(5):761--769.

\bibitem{Zhao2019DetectingTP}
Zhao J, et~al.
\newblock {Detecting time-evolving phenotypic topics via tensor factorization
  on electronic health records: Cardiovascular disease case study}.
\newblock Journal of Biomedical Informatics. 2019;98:103270.

\bibitem{Mullin2021LongitudinalKA}
Mullin S, et~al.
\newblock {Longitudinal K-means approaches to clustering and analyzing EHR
  opioid use trajectories for clinical subtypes}.
\newblock Journal of Biomedical Informatics. 2021;122:103889.

\bibitem{xu2022sepsis}
Xu Z, Mao C, Su C, Zhang H, Siempos I, Torres LK, et~al.
\newblock {Sepsis subphenotyping based on organ dysfunction trajectory}.
\newblock Critical Care. 2022;26(1):197.

\bibitem{Yang2022MachineLA}
Yang S, Varghese P, Stephenson E, Tu K, Gronsbell JL.
\newblock {Machine learning approaches for electronic health records
  phenotyping: A methodical review}.
\newblock Journal of the American Medical Informatics Association: JAMIA.
  2023;30(2):367--381.

\bibitem{He2023TrendsAO}
He T, et~al.
\newblock {Trends and opportunities in computable clinical phenotyping: A
  scoping review}.
\newblock Journal of Biomedical Informatics. 2023;140:104335.

\bibitem{Wang2019UnsupervisedML}
Wang Y, et~al.
\newblock {Unsupervised Machine Learning for the Discovery of Latent Disease
  Clusters and Patient Subgroups Using Electronic Health Records}.
\newblock Journal of biomedical informatics. 2020;Article no. 103364.

\bibitem{Ibrahim2019OnCS}
Ibrahim ZM, Wu H, Hamoud AA, Stappen L, Dobson RJB, Agarossi A.
\newblock {On classifying sepsis heterogeneity in the ICU: insight using
  machine learning}.
\newblock Journal of the American Medical Informatics Association : JAMIA.
  2020;27(3):437--443.

\bibitem{Oh2022UsingSC}
Oh W, et~al.
\newblock {Using sequence clustering to identify clinically relevant
  subphenotypes in patients with COVID-19 admitted to the intensive care unit.}
\newblock Journal of the American Medical Informatics Association: JAMIA.
  2022;29(3):489--499.

\bibitem{Xu2019IdentifyingSO}
Xu Z, et~al.
\newblock {Identifying sub-phenotypes of acute kidney injury using structured
  and unstructured electronic health record data with memory networks}.
\newblock Journal of biomedical informatics. 2020;102.
\newblock Article no. 103361.

\bibitem{Lasko2019ComputationalPD}
Lasko TA, Mesa DA.
\newblock {Computational Phenotype Discovery via Probabilistic Independence}.
\newblock arXiv:190711051. 2019;.

\bibitem{Smith2022OnlineCD}
Smith JO, Josef CS, Xie Y, Kamaleswaran R.
\newblock {Online Critical-State Detection of Sepsis Among ICU Patients using
  Jensen-Shannon Divergence}.
\newblock AMIA Annual Symposium. 2022;2022:982--991.

\bibitem{Estiri2021HighthroughputPW}
Estiri H, Strasser ZH, Murphy SN.
\newblock {High-throughput phenotyping with temporal sequences}.
\newblock J Am Medical Informatics Assoc. 2021;28:772--781.

\bibitem{Lee2020TemporalPU}
{Temporal Phenotyping using Deep Predictive Clustering of Disease Progression}.
  vol. 119; 2020.

\bibitem{Xu2005SurveyOC}
Xu R, Wunsch DC.
\newblock {Survey of clustering algorithms}.
\newblock IEEE Transactions on Neural Networks. 2005;16(3):645--678.

\bibitem{MacQueen1967SomeMF}
MacQueen J.
\newblock {Some methods for classification and analysis of multivariate
  observations}.
\newblock Proceedings of the Fifth Berkeley Symposium on Mathematical
  Statistics and Probability. 1967;1:281--297.

\bibitem{Johnson1967HierarchicalCS}
Johnson SC.
\newblock {Hierarchical clustering schemes}.
\newblock Psychometrika. 1967;32:241--254.

\bibitem{Ester1996ADA}
Ester M, Kriegel HP, Sander J, Xu X.
\newblock {A Density-Based Algorithm for Discovering Clusters in Large Spatial
  Databases with Noise}.
\newblock In: Proceedings of the Second International Conference on Knowledge
  Discovery and Data Mining; 1996. p. 226--231.

\bibitem{Bezdek1984FCMTF}
Bezdek JC, Ehrlich R, Full WE.
\newblock {FCM: The fuzzy c-means clustering algorithm}.
\newblock Computers \& Geosciences. 1984;10:191--203.

\bibitem{Cleuziou2008AnEV}
Cleuziou G.
\newblock {An extended version of the k-means method for overlapping
  clustering}.
\newblock In: 2008 19th International Conference on Pattern Recognition; 2008.
  p. 1--4.

\bibitem{Zhang1999KHarmonicM}
Zhang B, Hsu M, Dayal U.
\newblock {K-Harmonic Means - A Data Clustering Algorithm}.
\newblock Hewlett-Packard Laboratories; 1999. HPL-1999-124.

\bibitem{Ferraro2020SoftC}
Ferraro MB, Giordani P.
\newblock {Soft clustering}.
\newblock Wiley Interdisciplinary Reviews: Computational Statistics.
  2020;12(1).
\newblock Article no. e1480.

\bibitem{DRGcodes}
{Centers for Medicare \& Medicaid Services}. {MS-DRG Classifications and
  Software}; 2022.
\newblock Accessed December, 2022.
\newblock
  \url{https://www.cms.gov/Medicare/Medicare-Fee-for-Service-Payment/AcuteInpatientPPS/MS-DRG-Classifications-and-Software}.

\bibitem{ICDcodes}
{Centers for Disease Control}. {Classification of Diseases, Functioning, and
  Disability}; 2022.
\newblock Accessed October, 2022.
\newblock \url{http://www.cdc.gov/nchs/icd.htm}.

\bibitem{DRGtoICD}
{Centers for Medicare \& Medicaid Services}. {ICD-10-CM/PCS MS-DRG v36.0
  Definitions Manual}; 2018.
\newblock Accessed December, 2022.
\newblock
  \url{https://www.cms.gov/icd10m/version36-fullcode-cms/fullcode_cms/P0326.html}.

\bibitem{Otto2011TheLP}
Otto GP, et~al.
\newblock {The late phase of sepsis is characterized by an increased
  microbiological burden and death rate}.
\newblock Critical Care. 2011;15(4):R183.

\bibitem{Jain2013LowrankMC}
Jain P, Netrapalli P, Sanghavi S.
\newblock {Low-Rank Matrix Completion Using Alternating Minimization}.
\newblock In: Proceedings of the Forty-Fifth Annual ACM Symposium on Theory of
  Computing; 2013. p. 665--674.

\bibitem{Jiang23TimelineRE}
Jiang S, Han R, Chakrabarty K, Page D, Stead WW, Zhang AR.
\newblock {Timeline Registration for Electronic Health Records}.
\newblock In: AMIA 2023 Informatics Summit; 2023. .

\bibitem{Khanmohammadi2017AnIO}
Khanmohammadi S, Adibeig N, Shanehbandy S.
\newblock {An improved overlapping k-means clustering method for medical
  applications}.
\newblock Expert Syst Appl. 2017;67:12--18.

\bibitem{Basu2002SemisupervisedCB}
Basu S, Banerjee A, Mooney RJ.
\newblock {Semi-supervised Clustering by Seeding}.
\newblock In: Proceedings of the 19th International Conference on Machine
  Learning (ICML-2002); 2002. p. 19--26.

\bibitem{Park2009ASA}
Park HS, Jun CH.
\newblock {A simple and fast algorithm for K-medoids clustering}.
\newblock Expert Syst Appl. 2009;36(2):3336--3341.

\bibitem{Rousseeuw1987SilhouettesAG}
Rousseeuw PJ.
\newblock {Silhouettes: a graphical aid to the interpretation and validation of
  cluster analysis}.
\newblock Journal of Computational and Applied Mathematics. 1987;20:53--65.

\bibitem{Lipton2015LearningTD}
Lipton ZC, Kale DC, Elkan CP, Wetzel RC.
\newblock {Learning to Diagnose with LSTM Recurrent Neural Networks}.
\newblock ICLR. 2016;.

\bibitem{Paszke2019PyTorchAI}
Paszke A, et~al.
\newblock {PyTorch: An Imperative Style, High-Performance Deep Learning
  Library}.
\newblock In: Advances in Neural Information Processing Systems 32; 2019. p.
  8024--8035.

\bibitem{Pedregosa2011ScikitlearnML}
Pedregosa F, et~al.
\newblock {Scikit-learn: Machine Learning in Python}.
\newblock Journal of Machine Learning Research. 2011;12:2825--2830.

\bibitem{Gao2020DrAC}
Gao J, Xiao C, Glass L, Sun J.
\newblock {Dr. Agent: Clinical predictive model via mimicked second opinions}.
\newblock Journal of the American Medical Informatics Association: JAMIA.
  2020;27(7):1084--1091.

\end{thebibliography}

\clearpage
\newpage

\onecolumn
\appendix

\noindent \textbf{\Large Supplementary Materials for:}
\\\\
\noindent \textbf{\Large Soft Phenotyping for Sepsis via EHR Time-aware Soft Clustering}

\section{Algorithms of the proposed time-aware soft clustering}

\begin{algorithm}[hbt!]
\KwIn{$D = \{x_{i}, i=1,...,N\}, L = \{l_{i} \in \mathbb{R}^K, i=1,...,N\}, \beta_1, \beta_2, \eta, l_{rate}, K, t_{max}$}
\KwOut{$\{temp_{k}, k=1,...,K\}, M = \{\mu_{ik}\} \in \mathbb{R}^{N \times K}$}
\textbf{Step 1. Cluster centroids and cluster assignments initialization}\\
$\{temp_{k}, k=1,...,K\} \gets Initialization(D, L)$\;
\For{$x_{i} \in D$}{
    $m_{i}^{(0)} \gets Assign(x_{i}, \{temp_{k}, k=1,...,K\})$ \tcp{($Assign(\cdot)$ see Algorithm \ref{algo:2})}\
}
$t = 0$\;
\While{$t < t_{max}$}{
    \textbf{Step 2. Cluster centroids update}\\
    $t_{k}^{(0)}, s_{k}^{(0)} = 0, 0$\;
    \For{k=1:K}{
        \For{$x_{i} \in Cluster_{k}$}{
            $\alpha_{i}^{(t)} \gets \frac{1}{|m_{i}^{(t)}|^2}$\;
            $temp_{k}^{i} \gets |m_{i}^{(t)}| \cdot x_{i} - \sum_{c \in m_{i}^{(t)} \setminus {k}} temp_{c}^{(t)}$\;
            $t_{k}^{(t+1)} \gets t_{k}^{(t)} + \alpha_{i}^{(t)} \cdot temp_{k}^{i}$\;
            $s_{k}^{(t+1)} \gets s_{k}^{(t)} + \alpha_{i}^{(t)}$\;
        }
        $temp_{k}^{(t+1)} \gets \frac{t_{k}^{(t+1)}}{s_{k}^{(t+1)}}$\;
    }
    \textbf{Step 3. Cluster centroids calibration}\\
    $\{temp_{k}^{(t+1)}\} \gets totLossOpt(D, \{temp_{k}^{(t+1)}\}, \{m_i^{(t)}\}, L, \beta_1, \beta_2, \eta, l_{rate})$ \tcp{($totLossOpt(\cdot)$ see Algorithm \ref{algo:3})}\
    \textbf{Step 4. Cluster assignments update}\\
    \For{$x_{i} \in D$}{
        $m_{i}^{(t+1)} \gets Assign(x_{i}, \{temp_{k}^{(t+1)}, k=1,...,K\}, m_{i}^{(t)})$\;
    }
}
\textbf{Step 5. Compute final membership}\\
$\{d_{ik}\} \in \mathbb{R}^{N \times K} \gets ComputeDistMat(\{x_i, i=1, ..., N\}, \{temp_{k}^{(t+1)}, k=1,...,K\})$\;
$\{\mu_{ik}\} \in \mathbb{R}^{N \times K} \gets \frac{d_{ik}^{-3}}{\sum_{j=1}^K d_{ij}^{-3}}$\;
\caption{Time-aware Soft Clustering}
\label{algo:1}
\end{algorithm}

\begin{algorithm}[hbt!]
\KwIn{$D = \{x_{i}, i=1,...,N\}, \{temp_{k}, k=1,...,K\}, \{m_{i}^{old}, i=1,...,N\}$}
\KwOut{$\{m_{i}, i=1,...,N\}$}
\For{$x_{i} \in D$}{
    \textbf{Step 1.}\\
    \For{k=1:K}{
        $d_{i,k} \gets Distance(x_{i}, temp_{k})$\;
    }
    $d_{i,min} \gets \min\{d_{i,k}, k=1,...,K\}$\;
    $m_{i} \gets \{\} \cup c \leftarrow \min\{d_{i,k}, k=1,...,K\}$\;
    $\Phi(x_{i}) \gets \frac{\sum_{c \in m_{i}} temp_{c}}{|m_{i}|}$\;
    \textbf{Step 2.}\\
    $m_{i}^{'} \gets m_{i} \cup c \leftarrow \min\{d_{i,k}, k \notin m_{i}\}$\;
    $\Phi(x_{i})^{'} \gets \frac{\sum_{c \in m_{i}^{'}} temp_{c}}{|m_{i}^{'}|}$\;
    \textbf{Step 3.}\\
    \If{$\lVert x_{i} - \Phi(x_{i})^{'} \rVert < \lVert x_{i} - \Phi(x_{i}) \rVert$}{
        $m_{i} \gets m_{i}^{'}$\;
        $\Phi(x_{i}) \gets \Phi(x_{i})^{'}$\;
        Go to \textbf{Step 2.}\\
    }
    \Else{
        $\Phi(x_{i})^{old} \gets compute(m_{i}^{old})$\;
        \If{$\lVert x_{i} - \Phi(x_{i}) \rVert \leq \lVert x_{i} - \Phi(x_{i})^{old} \rVert$}{
            \textbf{Return} $m_{i}$\;
        }
        \Else{
            \textbf{Return} $m_{i}^{old}$\;
        }
    }
}
\caption{Cluster Assignments}
\label{algo:2}
\end{algorithm}

\begin{algorithm}[hbt!]
\KwIn{$D = \{x_{i}, i=1,...,N\}, \{temp_{k}, k=1,...,K\}, \{m_{i}, i=1,...,N\}, L = \{l_{i} \in \mathbb{R}^K, i=1,...,N\}, \beta_1, \beta_2, \eta, l_{rate}$}
\KwOut{$\{temp_{k}^{calib}, k=1,...,K\}$}
\For{$x_i \in D$}{
    \For{$k = 1:K$}{
        $u_{ik} \gets \frac{1}{(\frac{Distance(x_i, temp_{k})}{\sum_{j=1}^K Distance(x_i, temp_{j})})^{\frac{2}{\eta-1}}}$\;
    }
}
$unsupLoss \gets \sum_{i=1}^N \sum_{k=1}^K u_{ik}^{\eta} Distance(x_i, temp_{k})$\;
\For{$x_i \in D \land sum(l_i)==1$}{
    \tcp{choose subjects with a single ICD group label}\
    $tLoss \gets Distance(x_i, temp_{l \in l_i})$\;
    $ntLoss \gets Distance(x_i, \{temp_{l \notin l_i})\}$\;
}
$supLoss \gets \beta_1 \cdot tLoss - \beta_2 \cdot ntLoss $\;
$totLoss \gets unsupLoss + supLoss$\;
$\{temp_{k}^{calib}\} \gets \{temp_{k}\} - l_{rate} \cdot \nabla \{temp_{k}\}$\;
\caption{Cluster centroids calibration}
\label{algo:3}
\end{algorithm}

\newpage
\section{Additional results and discussion for early-warning prediction}

\begin{figure}[hbt!]
    \centering
    \subfigure[MIMIC-IV]{
    \includegraphics[width=0.5\linewidth]{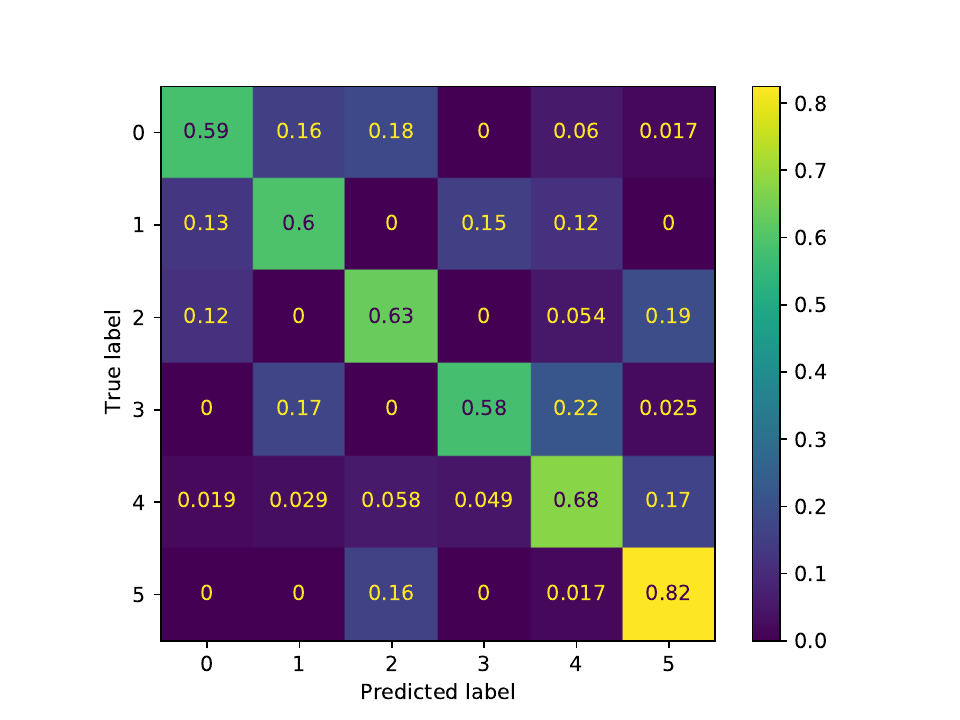}
    \label{fig:5}
    }
    \subfigure[eICU]{
	\includegraphics[width=0.5\linewidth]{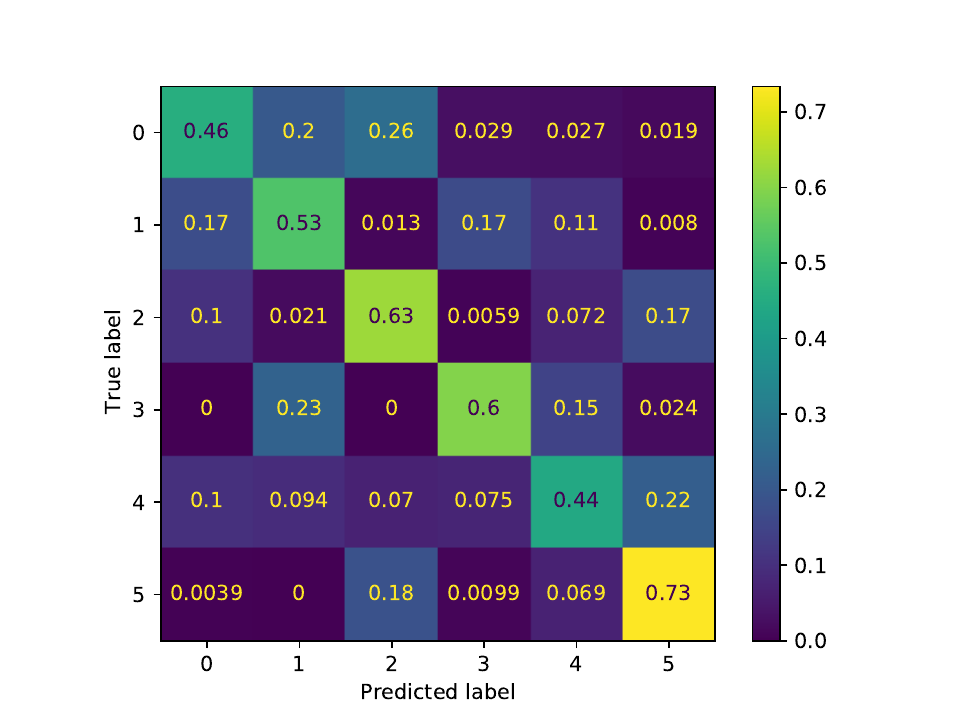}
    \label{fig:6}
    }
    \caption{Normalized confusion matrices over the ground truth for sepsis early-warning prediction using the MIMIC-IV and the eICU datasets.}
\end{figure}

We further present the corresponding confusion matrices for the early-warning prediction using the MIMIC-IV and the eICU datasets in Fig. \ref{fig:5} and Fig. \ref{fig:6}, respectively. Classes 0-5 in the figure correspond to the hybrid sub-phenotypes 1-6. Each value in the matrix represents the normalized value over the ground truth.

We observe that the model provides the strongest predictions to hybrid sub-phenotype 6 patients from both datasets. In addition, the model has strong prediction performance for hybrid sub-phenotype 5 patients in the MIMIC-IV dataset and hybrid sub-phenotypes 3 and 4 patients in the eICU dataset. The prediction results may provide insights for treatments/medications at a later stage.

\section{Additional figures}

\begin{figure*}[!hbt]
  \centering
  \captionsetup{justification=centering}
  \includegraphics[width=\textwidth]{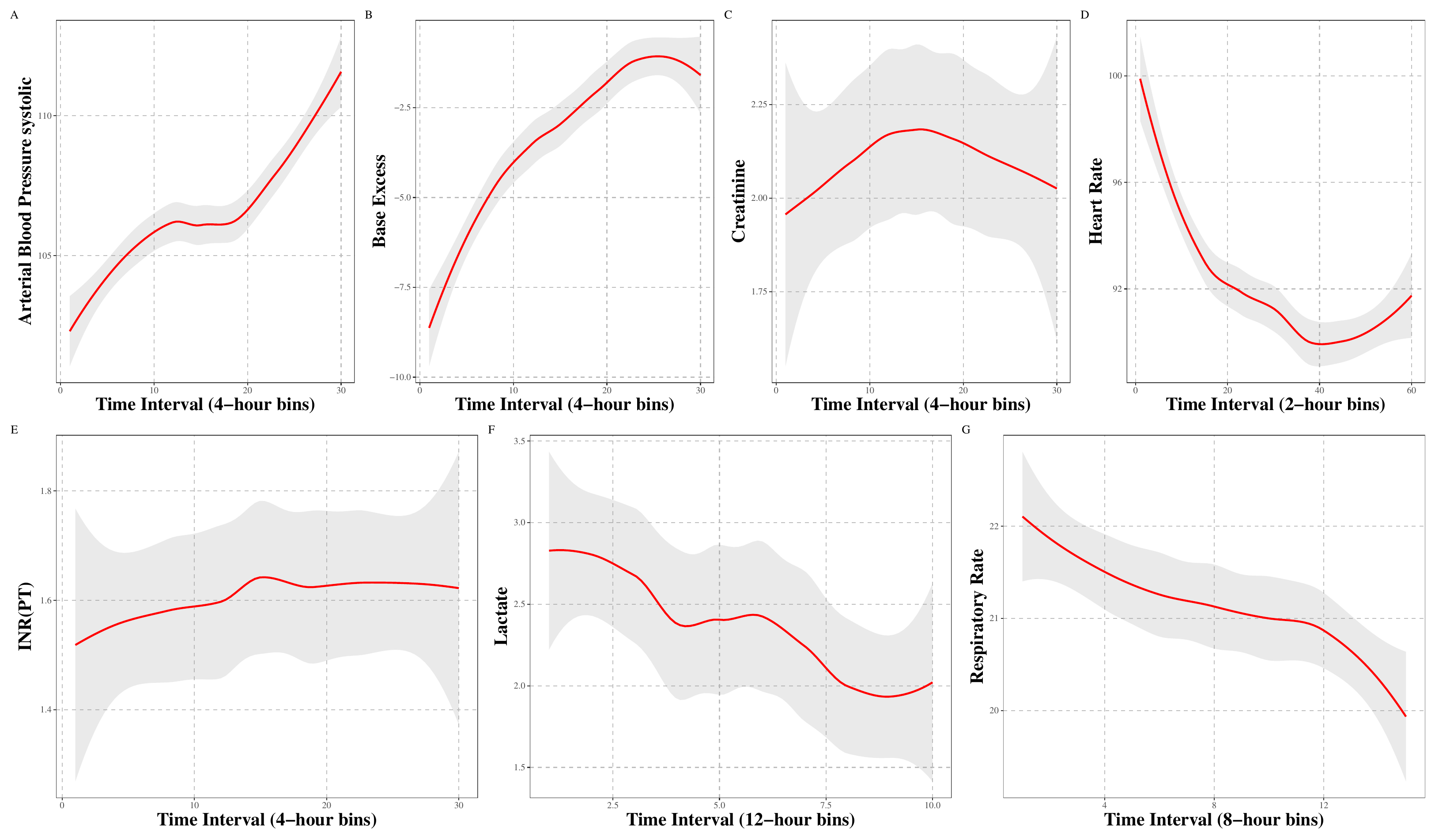}
  \caption{Feature values of hybrid sub-phenotype 1 using the MIMIC-IV dataset.}
  \label{fig:s3}
\end{figure*}

\begin{figure*}[!hbt]
  \centering
  \captionsetup{justification=centering}
  \includegraphics[width=\textwidth]{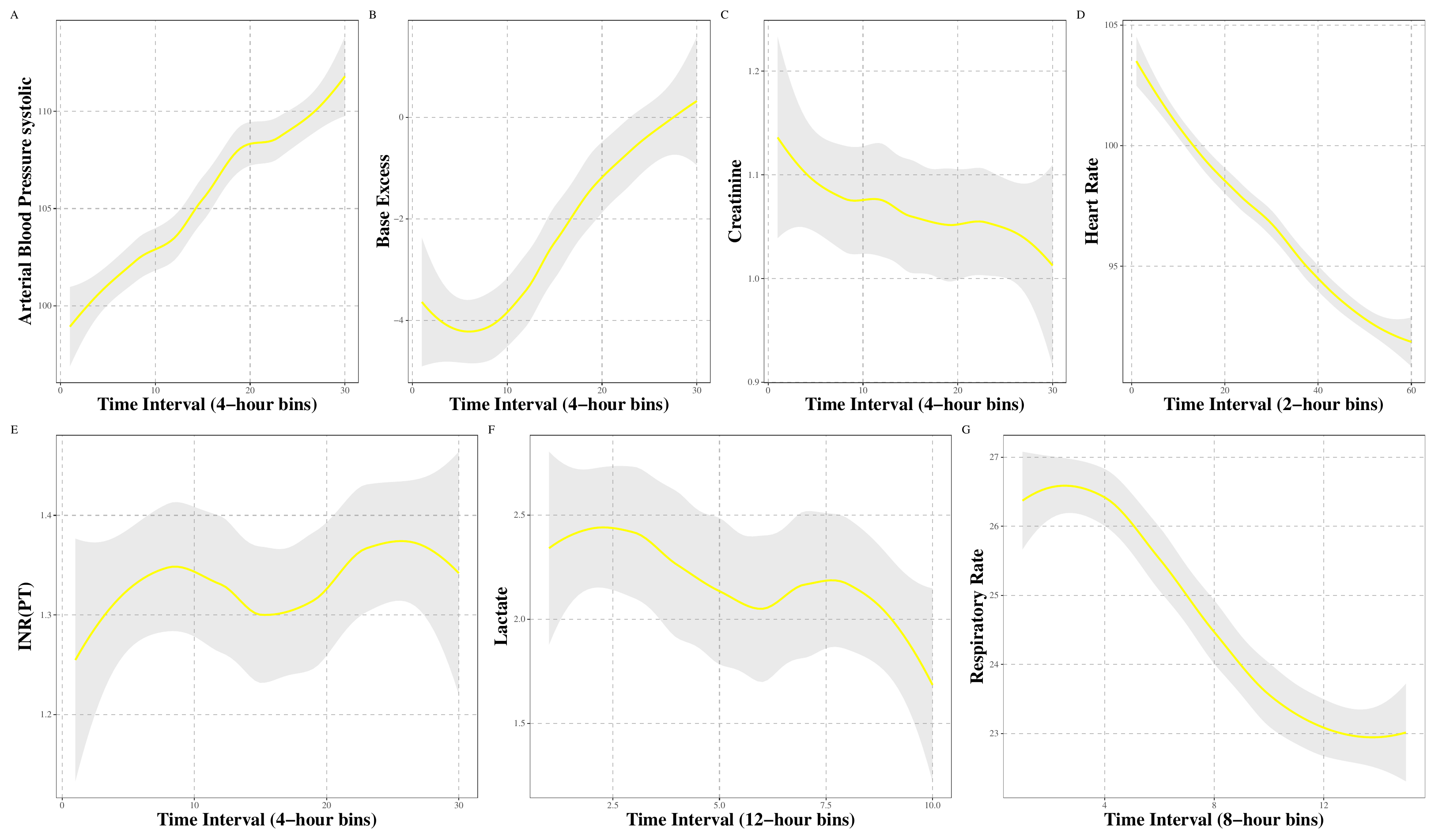}
  \caption{Feature values of hybrid sub-phenotype 2 using the MIMIC-IV dataset.}
  \label{fig:s4}
\end{figure*}

\begin{figure*}[!hbt]
  \centering
  \captionsetup{justification=centering}
  \includegraphics[width=\textwidth]{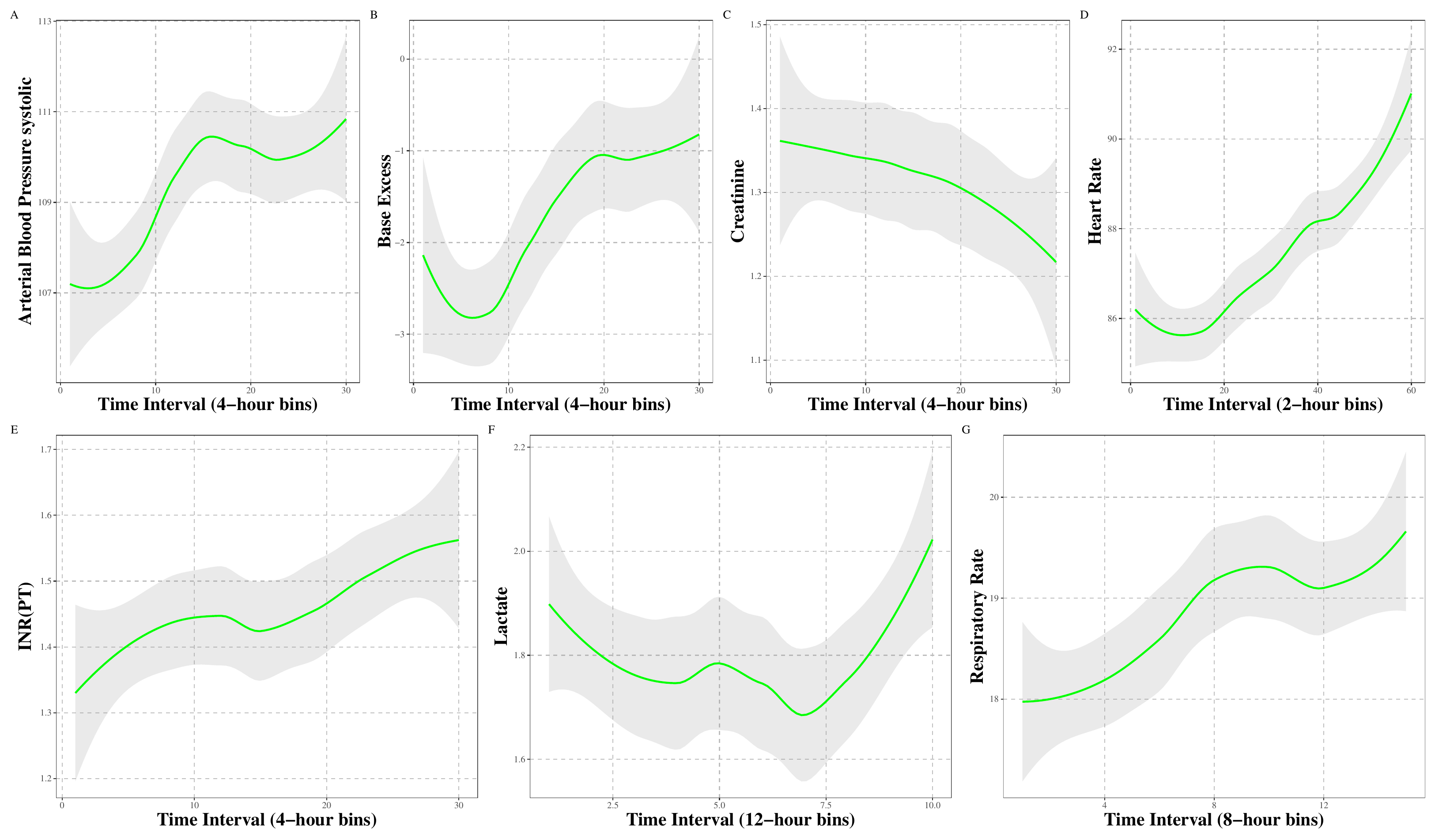}
  \caption{Feature values of hybrid sub-phenotype 3 using the MIMIC-IV dataset.}
  \label{fig:s5}
\end{figure*}

\begin{figure*}[!hbt]
  \centering
  \captionsetup{justification=centering}
  \includegraphics[width=\textwidth]{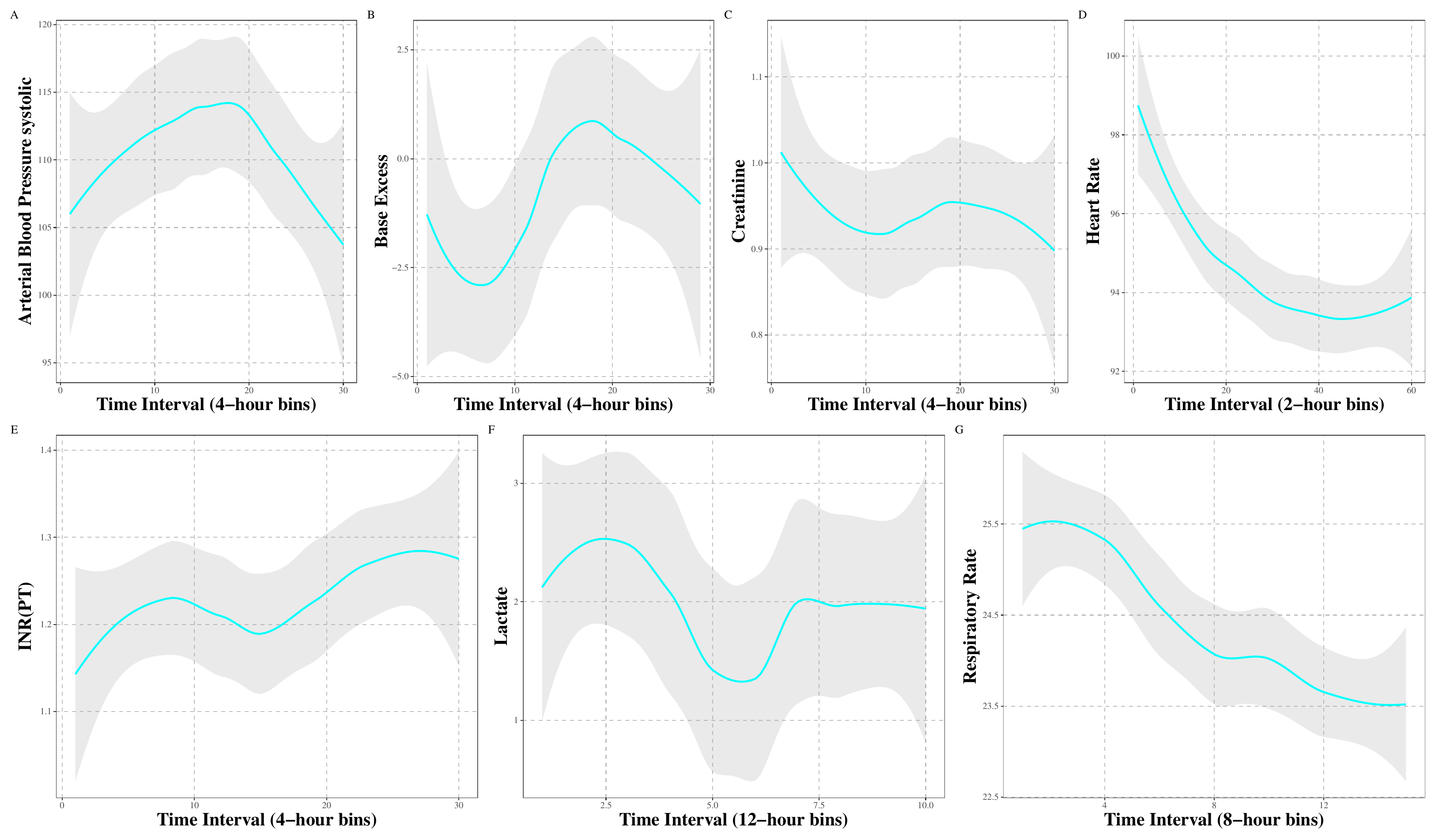}
  \caption{Feature values of hybrid sub-phenotype 4 using the MIMIC-IV dataset.}
  \label{fig:s6}
\end{figure*}

\begin{figure*}[!hbt]
  \centering
  \captionsetup{justification=centering}
  \includegraphics[width=\textwidth]{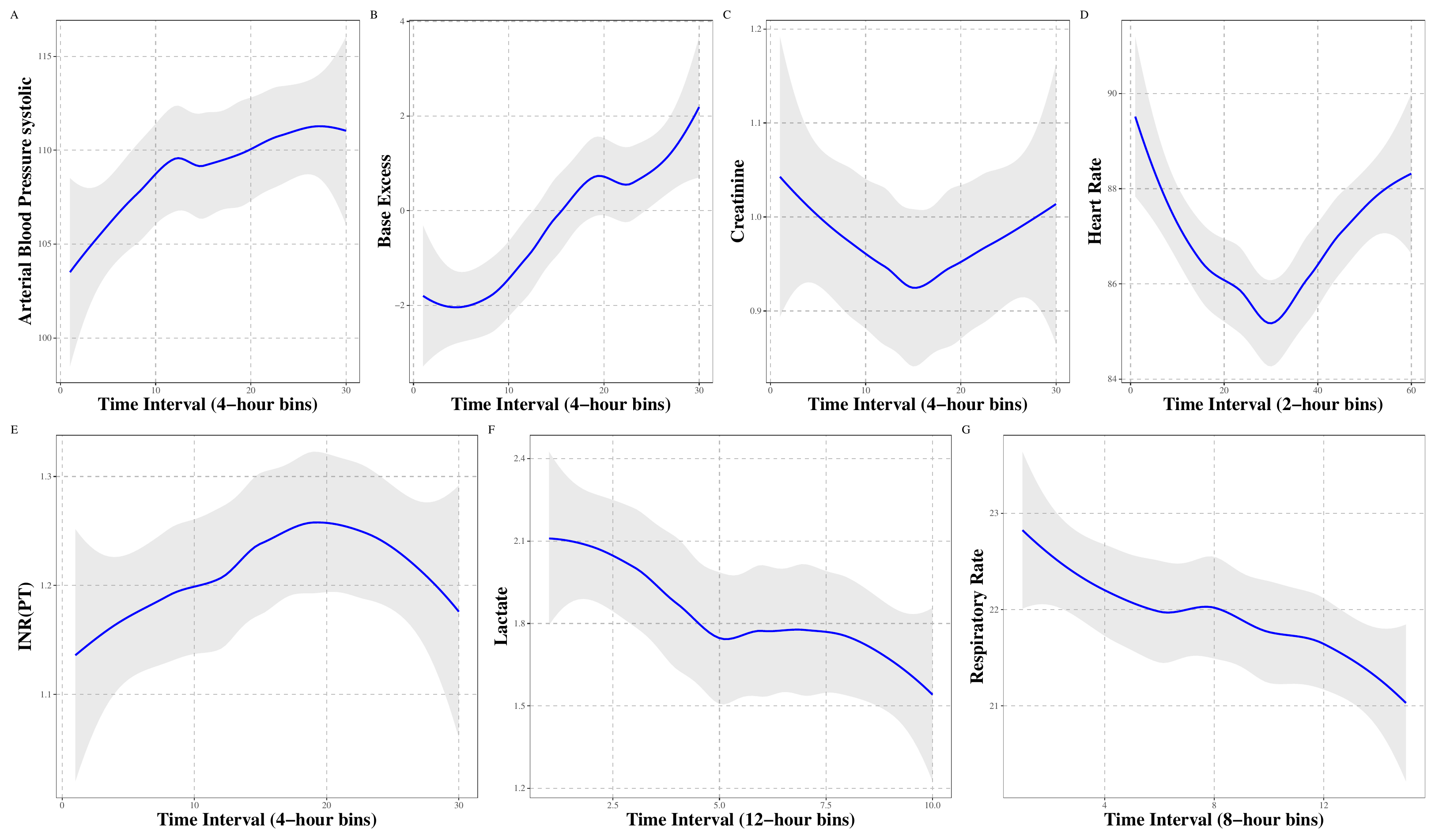}
  \caption{Feature values of hybrid sub-phenotype 5 using the MIMIC-IV dataset.}
  \label{fig:s7}
\end{figure*}

\begin{figure*}[!hbt]
  \centering
  \captionsetup{justification=centering}
  \includegraphics[width=\textwidth]{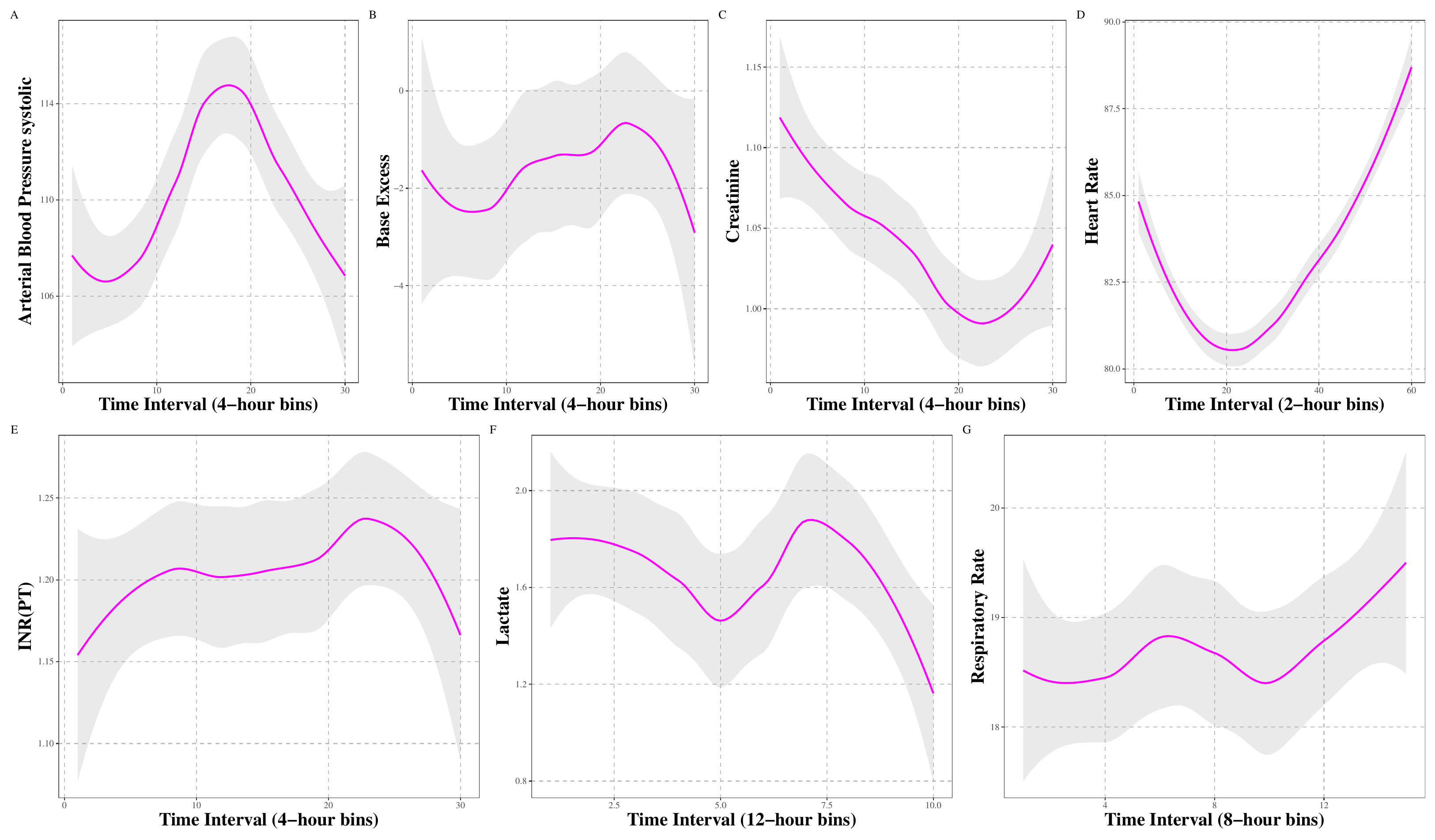}
  \caption{Feature values of hybrid sub-phenotype 6 using the MIMIC-IV dataset.}
  \label{fig:s8}
\end{figure*}

\begin{figure*}[!hbt]
  \centering
  \captionsetup{justification=centering}
  \includegraphics[width=.95\textwidth]{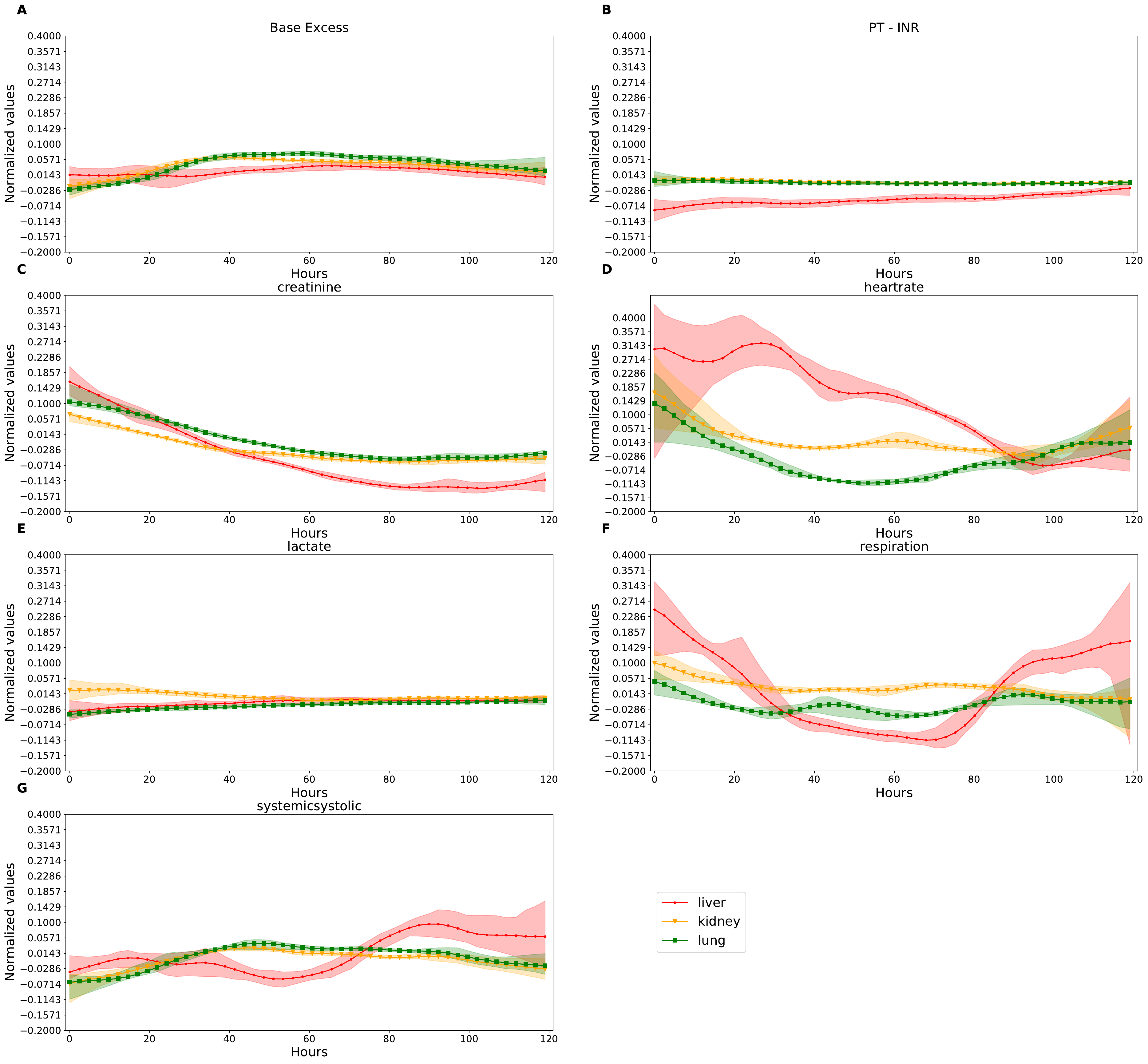}
  \caption{Soft clustering centroids per feature after smoothing obtained from the eICU dataset. The red centroid is initialized with the liver dysfunction type; the yellow centroid with the kidney dysfunction type; and the green centroid with the lung dysfunction type.}
  \label{fig:s1}
\end{figure*}

\begin{figure}[!hbt]
  \centering
  \captionsetup{justification=centering}
  \includegraphics[width=.6\linewidth]{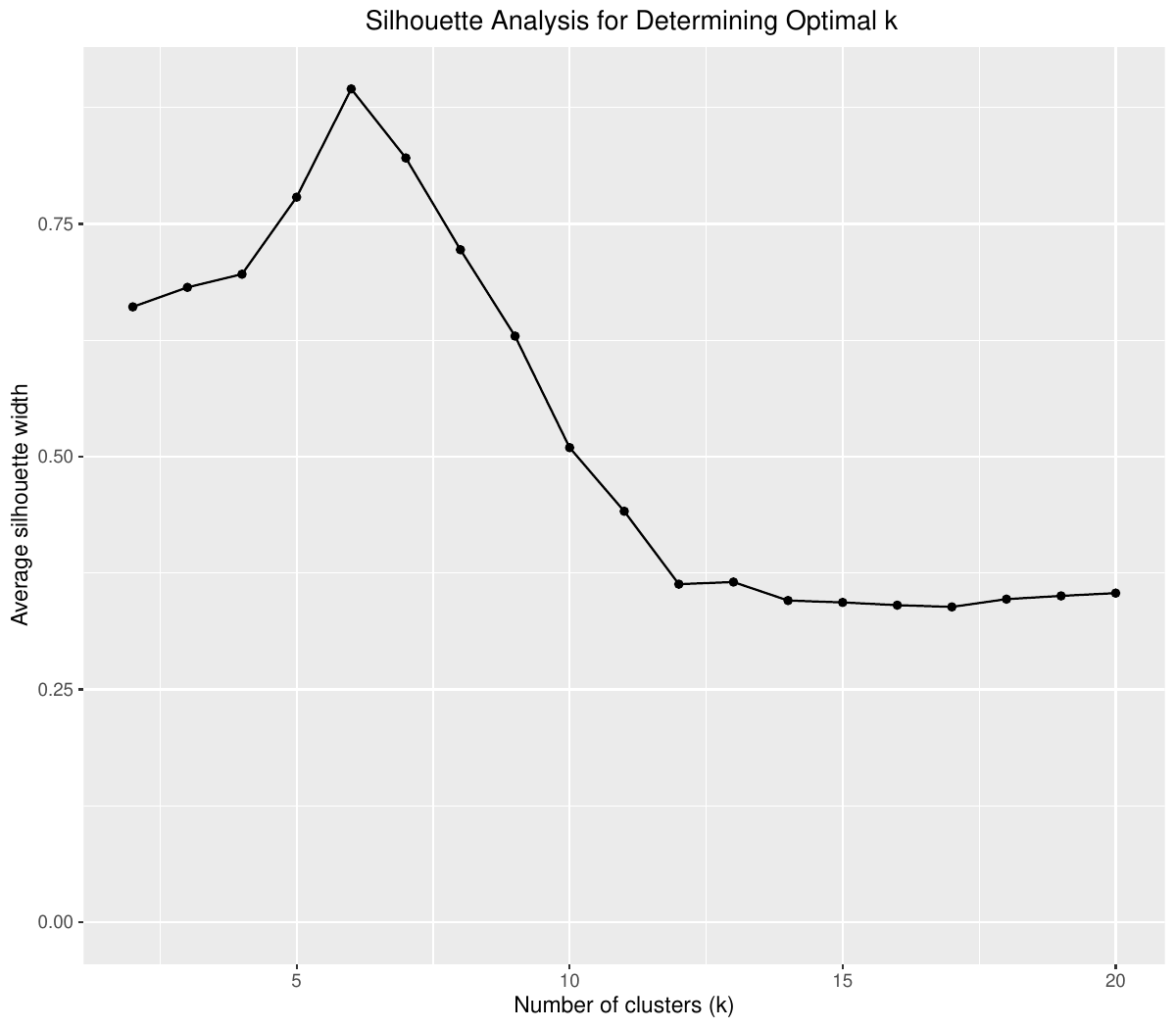}
  \caption{Cluster number selection for K-Medoids clustering using the eICU dataset.}
  \label{fig:s2}
\end{figure}

\begin{figure*}[!hbt]
  \centering
  \captionsetup{justification=centering}
  \includegraphics[width=\textwidth]{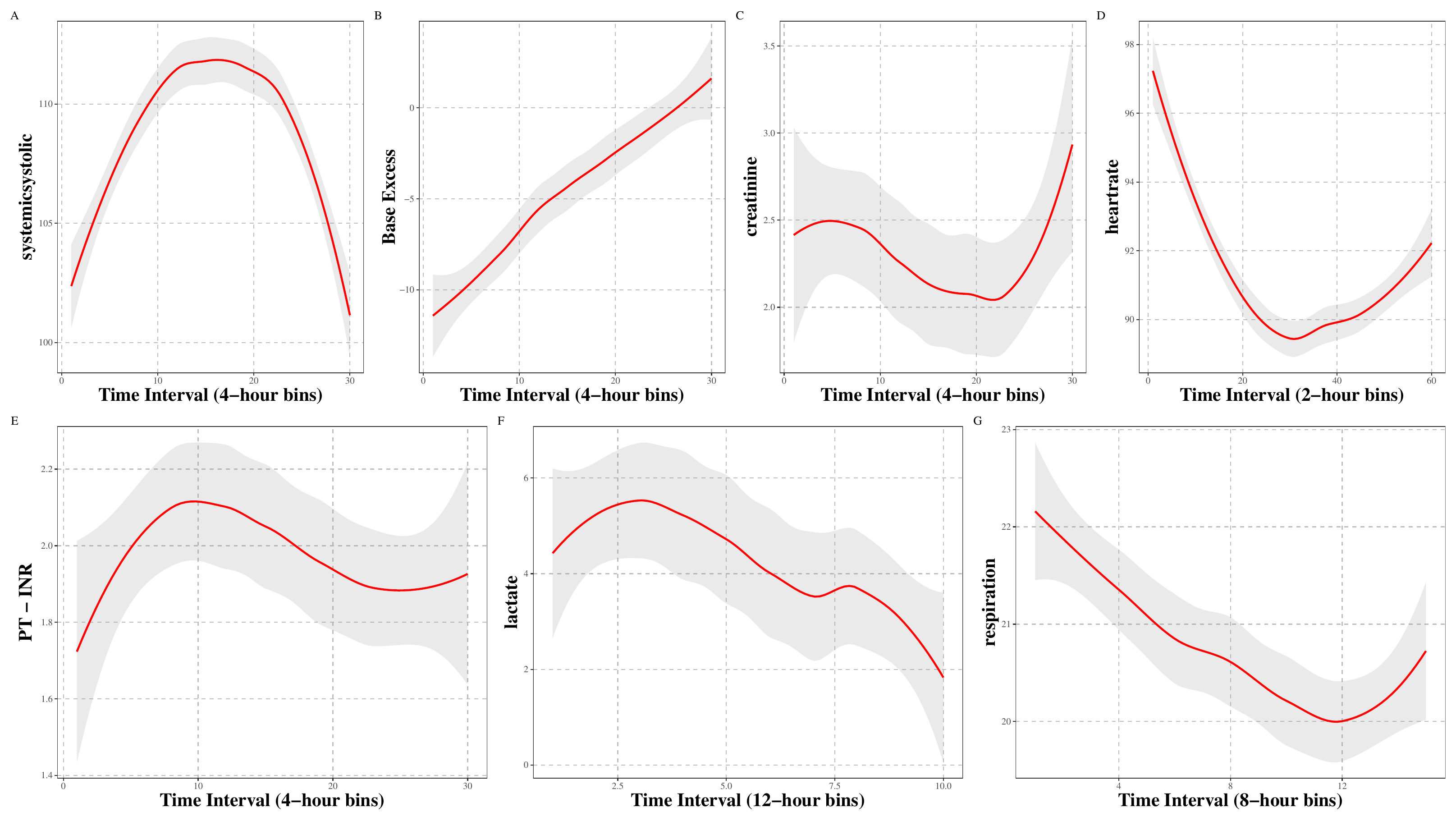}
  \caption{Feature values of hybrid sub-phenotype 1 using the eICU dataset.}
  \label{fig:s9}
\end{figure*}

\begin{figure*}[!hbt]
  \centering
  \captionsetup{justification=centering}
  \includegraphics[width=\textwidth]{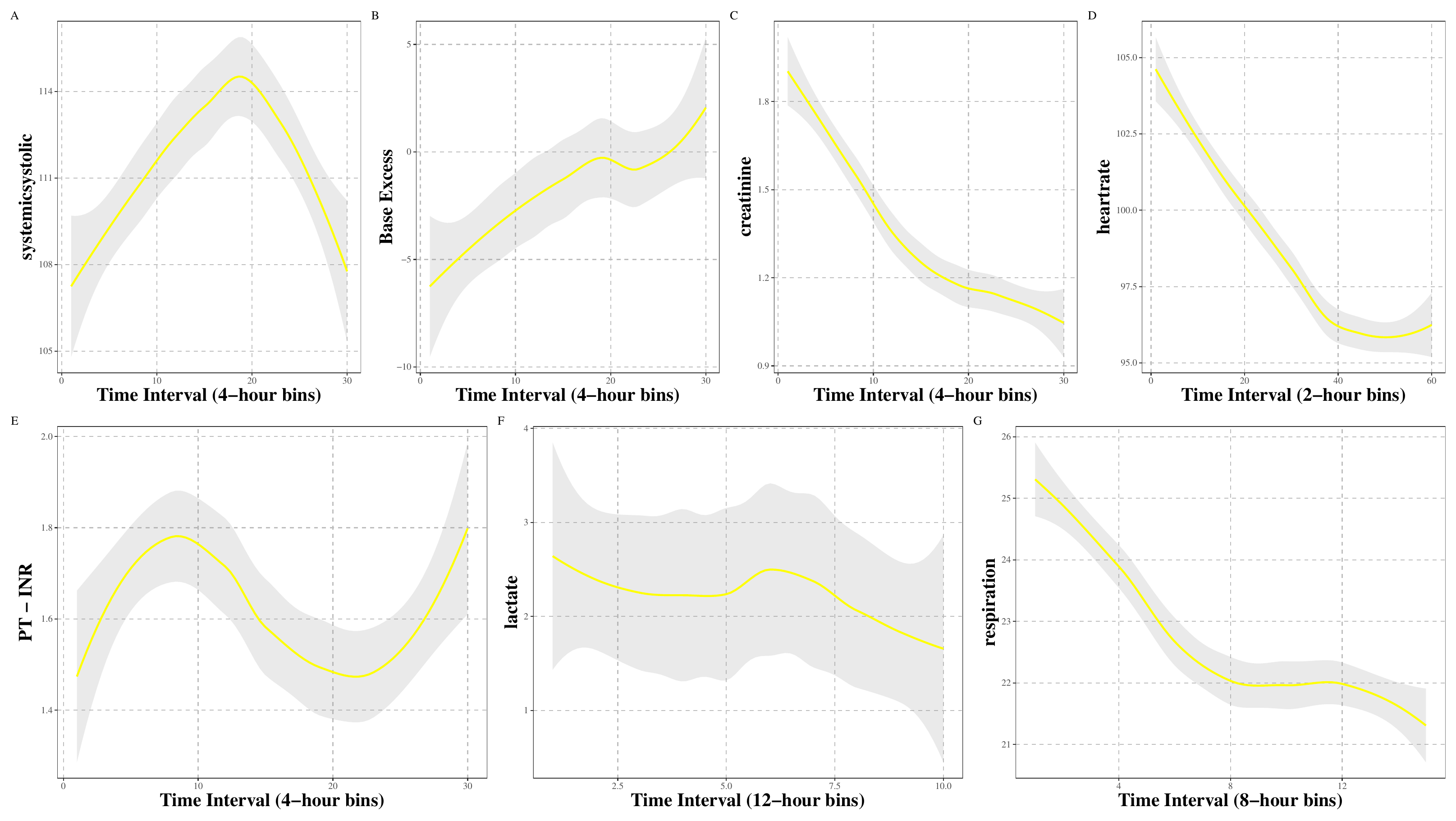}
  \caption{Feature values of hybrid sub-phenotype 2 using the eICU dataset.}
  \label{fig:s10}
\end{figure*}

\begin{figure*}[!hbt]
  \centering
  \captionsetup{justification=centering}
  \includegraphics[width=\textwidth]{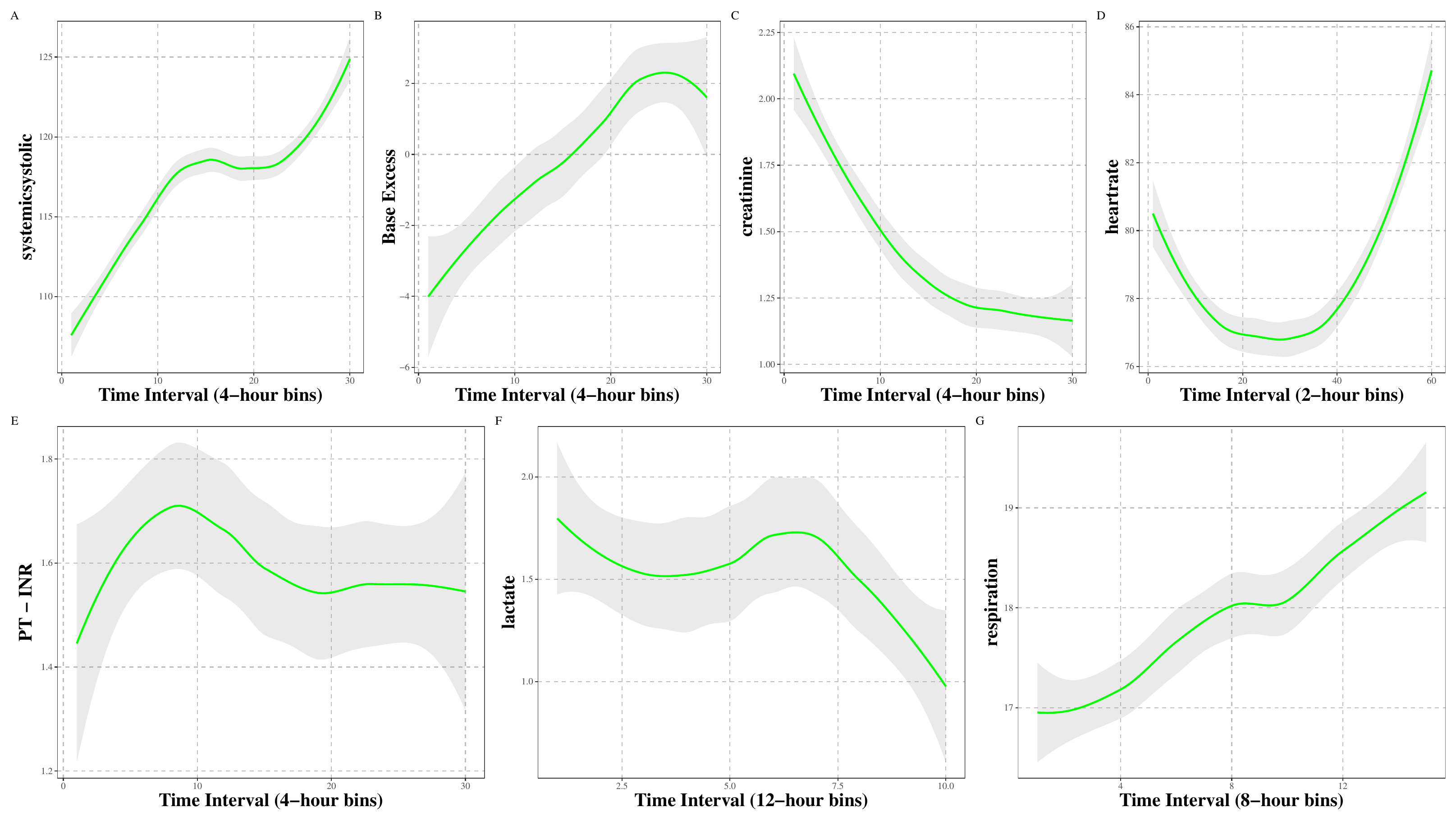}
  \caption{Feature values of hybrid sub-phenotype 3 using the eICU dataset.}
  \label{fig:s11}
\end{figure*}

\begin{figure*}[!hbt]
  \centering
  \captionsetup{justification=centering}
  \includegraphics[width=\textwidth]{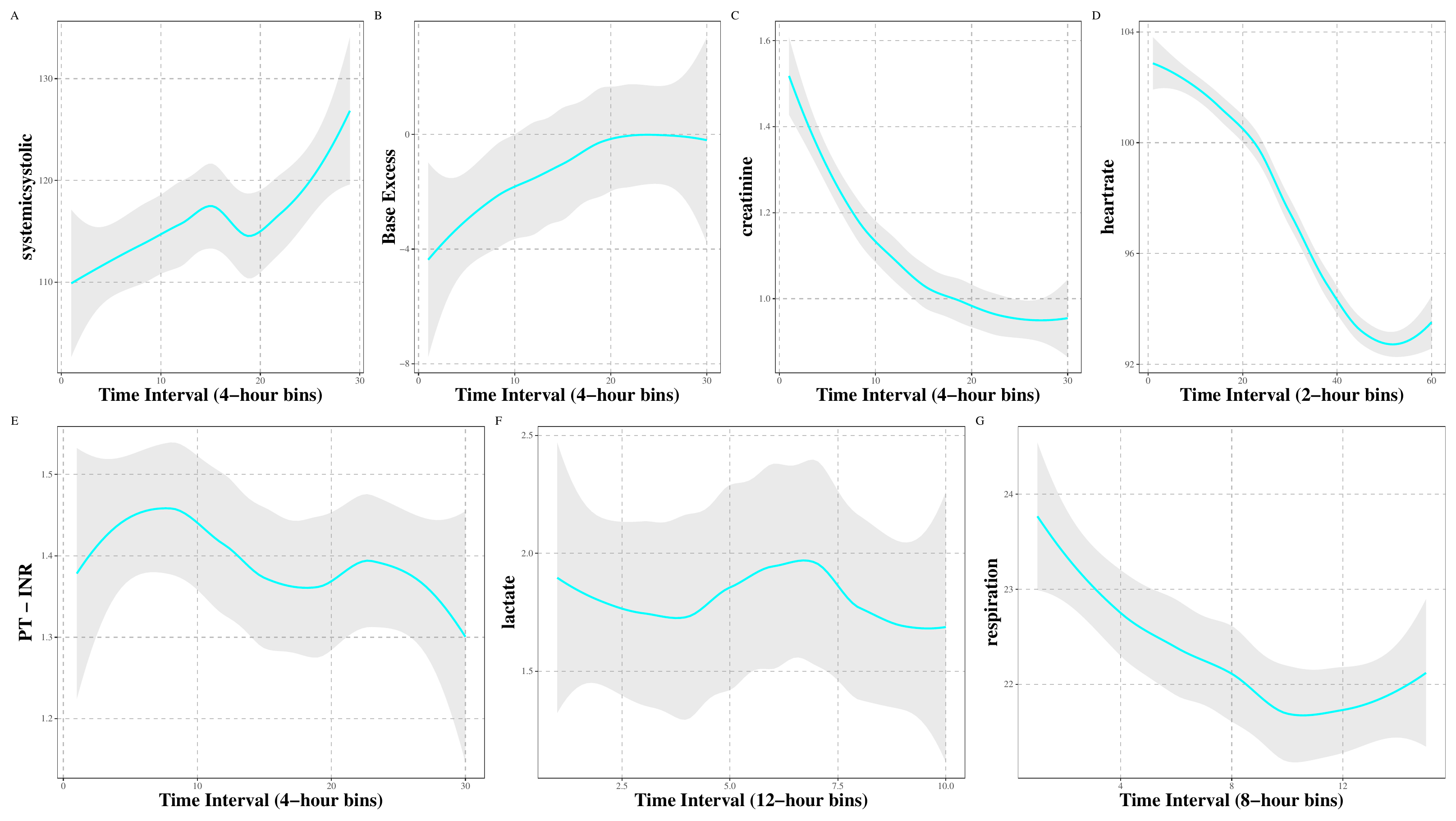}
  \caption{Feature values of hybrid sub-phenotype 4 using the eICU dataset.}
  \label{fig:s12}
\end{figure*}

\begin{figure*}[!hbt]
  \centering
  \captionsetup{justification=centering}
  \includegraphics[width=\textwidth]{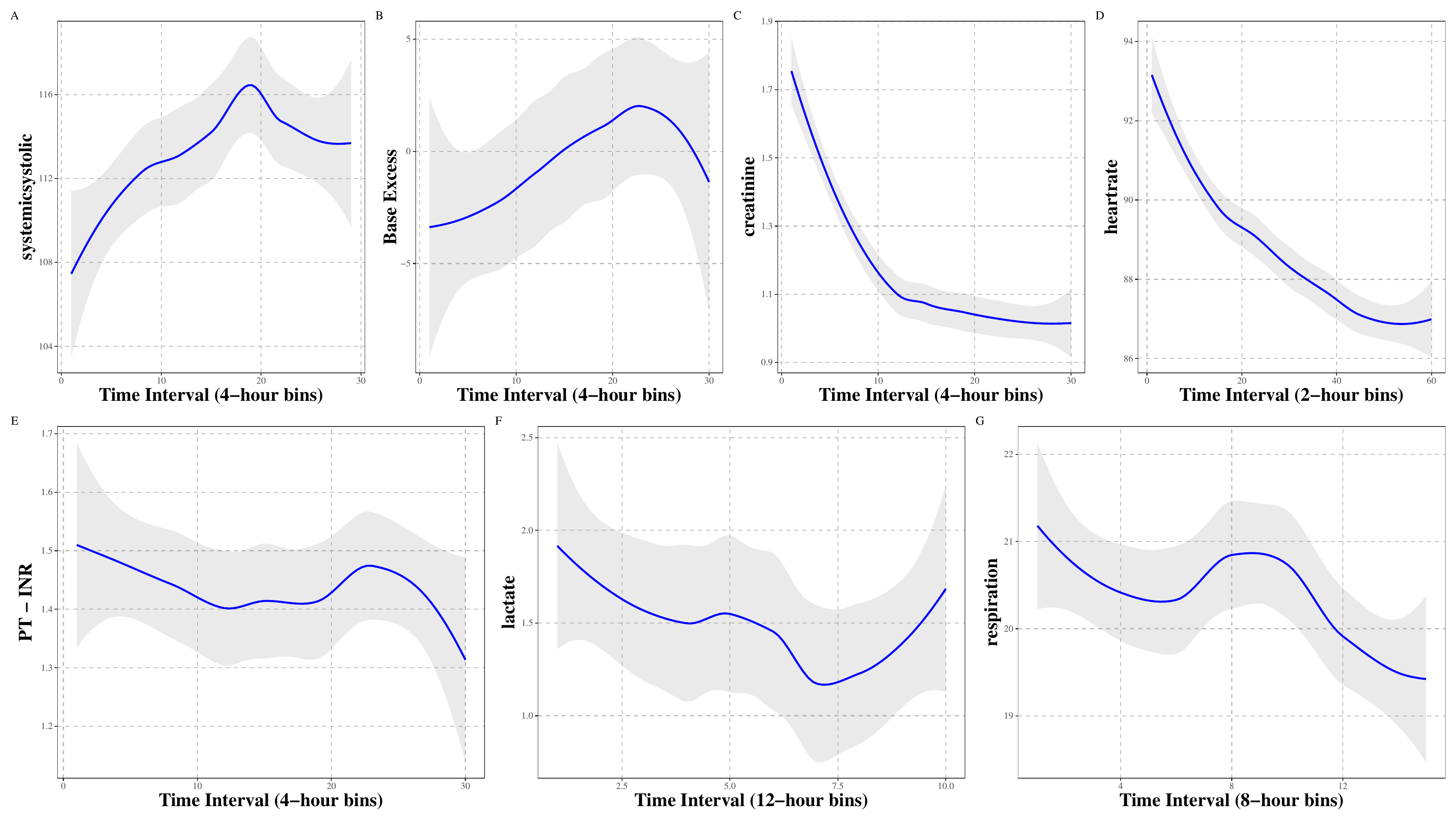}
  \caption{Feature values of hybrid sub-phenotype 5 using the eICU dataset.}
  \label{fig:s13}
\end{figure*}

\begin{figure*}[!hbt]
  \centering
  \captionsetup{justification=centering}
  \includegraphics[width=\textwidth]{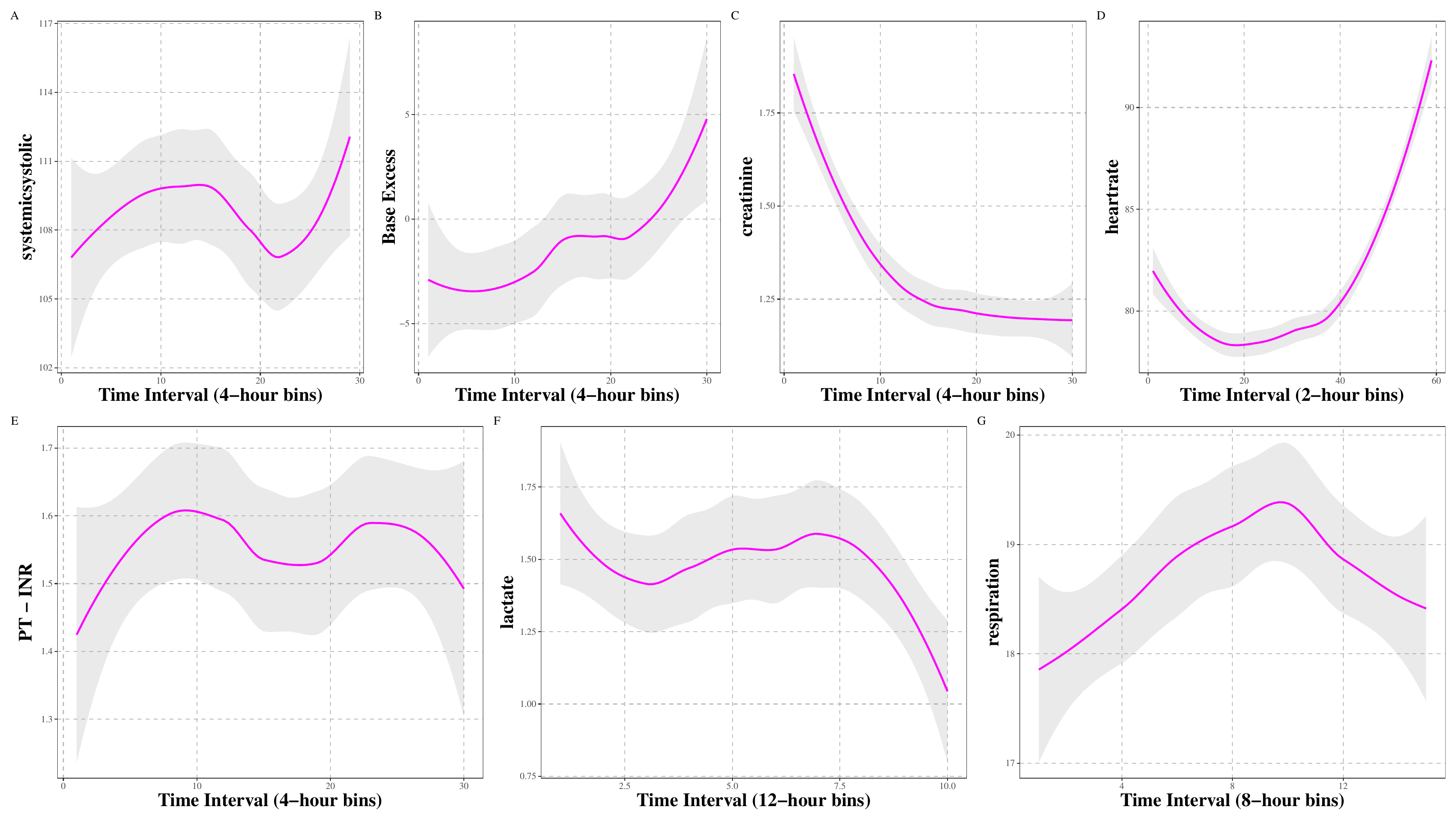}
  \caption{Feature values of hybrid sub-phenotype 6 using the eICU dataset.}
  \label{fig:s14}
\end{figure*}

\end{document}